\documentclass{article}
\usepackage{cite,comment}
\usepackage{amsmath,amssymb,amsfonts}
\usepackage{graphicx}
\usepackage{textcomp}
\usepackage{xcolor}
\usepackage{bbm}
\usepackage{amsthm,amsmath,fullpage,setspace,amsfonts,amsthm,placeins}
\usepackage{graphicx}
\usepackage{mathrsfs}
\usepackage{amsmath}
\usepackage{xcolor}
\usepackage{arydshln}
\usepackage{subfig}
\usepackage{algorithm}
\usepackage{xspace}
\usepackage{hyperref}
\usepackage[export]{adjustbox}

\usepackage[noend]{algpseudocode}

\newcommand{\alg}{\textsc{TASBM}\xspace}

\newcommand{\hide}[1]{}
\newcommand{\xhdr}[1]{\vspace{1.7mm}\noindent{{\bf #1.}}}

\newtheorem{theorem}{Theorem}

\newtheorem{lemma}{Lemma}

\newcommand{\ie}{\emph{i.e.}}

\newcommand{\E}{\mathbb{E}}
\newcommand{\bm}{M}

\newtheorem{definition}{\textbf{Definition}}
\graphicspath{{./FIG/}}

\begin{document}

%\title{Embedded Index Coding}

%\author{Alexandra Porter and Mary Wootters\thanks{AP is with the Department of Computer Science, Stanford University.  MW is with the Departments of Computer Science and Electrical Engineering, Stanford University.  This work is partially supported by NSF grant CCF-1657049 and NSF CAREER grant CCF-1844628.}}

\title{Analytical Models for Motifs in Temporal Networks: \\Discovering Trends and Anomalies}
\iffalse\author{
William L. Hamilton\\
\texttt{wleif@stanford.edu}
\and
Rex Ying\\
\texttt{rexying@stanford.edu}
\and
Jure Leskovec \\
\texttt{jure@cs.stanford.edu}
\vspace{10pt}
\sharedaffiliation
Department of Computer Science\\
Stanford University\\
Stanford, CA, 94305
}\fi

\author{Alexandra Porter\\
\texttt{amporter@cs.stanford.edu}\\
 Department of Computer Science\\
Stanford University\\
Stanford, CA 
\and  Baharan Mirzasoleiman\\
\texttt{baharan@cs.ucla.edu}\\
Department of Computer Science\\
University of California Los Angeles\\
Los Angeles, CA
 \and 
Jure Leskovec\\
\texttt{jure@cs.stanford.edu}\\
 Department of Computer Science\\
Stanford University\\
Stanford, CA } %\thanks{AP is with the Department of Computer Science, Stanford University.  AP is partially supported by the National Science Foundation Graduate Research Fellowship under Grant No. DGE-1656518}}
\maketitle

\begin{abstract}
% !TEX root =motifs_newarxiv.tex

%% Jure
Dynamic evolving networks capture temporal relations in domains such as social networks, communication networks, and financial transaction networks. 
In such networks, temporal motifs, which are repeated sequences of time-stamped edges/transactions, offer valuable information about the networks' evolution and function.
However, currently no analytical models for temporal graphs exist and there are no models that would allow for scalable modeling of temporal motif frequencies over time.
% to spot trends and anomalies statistically significant temporal motifs have to be identified. Due to high computational complexity, this is infeasible using existing approaches.
%
Here, we develop the {\em Temporal Activity State Block Model (TASBM)}, to model temporal motifs in temporal graphs. We develop efficient model fitting methods and derive closed-form expressions for the expected motif frequencies and their variances in a given temporal network, thus enabling the discovery of statistically significant temporal motifs. 
Our TASMB framework %is accurate 
can accurately track the changes in the expected motif frequencies over time,
and also scales well to networks with tens of millions of edges/transactions as it does not require time-consuming generation of many random temporal networks and then computing motif counts for each one of them.
%In contrast to current approaches, which require generating many random networks and then computing motif counts for each of them, our TASMB model directly provides expectations and variances of motif counts and
%our framework does not require generating randomized network ensembles, and thus scales to large temporal networks. Experiments show that our model accurately tracks motif frequencies while the number of nodes and number of edges change over time.
We show that TASBM is able to model changes in temporal activity over time in a network of financial transactions, a phone call, and an email network. Additionally, we show that deviations from the expected motif counts calculated by our analytical framework correspond to anomalies in the financial transactions and phone call networks.

\end{abstract}

\section{Introduction}
\label{sec:intro}
% !TEX root = motifs_newarxiv.tex

 \begin{figure}[t]
	\centering
	\subfloat[{}]{\includegraphics[width=.4\textwidth]{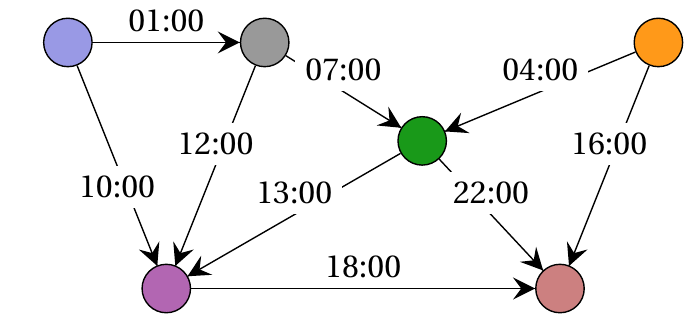}}
	\subfloat[{\label{fig:1b}}]{\includegraphics[width=.22\textwidth]{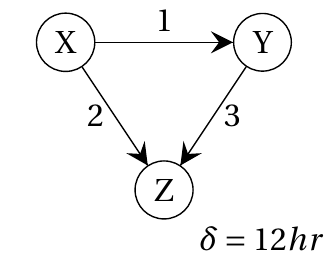}}\\
	\subfloat[{}]{\includegraphics[width=.18\textwidth]{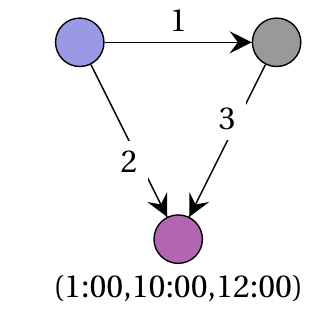}
		\includegraphics[width=.2\textwidth]{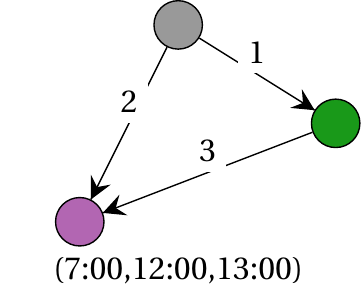}} 
	\subfloat[{}]{\includegraphics[width=.2\textwidth]{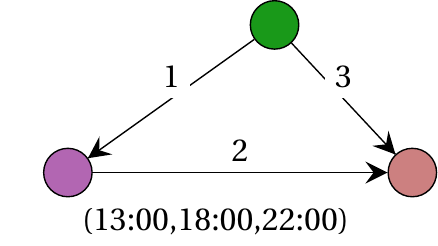}
		\includegraphics[width=.18\textwidth]{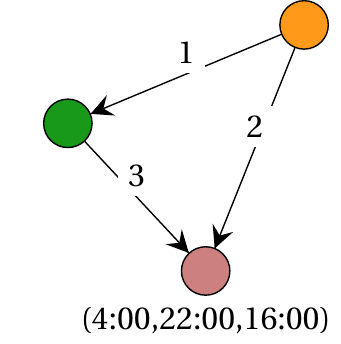}}
	\vspace{-0.2cm}
	\caption{(a) A temporal network with edges appearing over a day, (b) a motif $M$ with a temporal window of 12 hours, (c) examples of $M$ in the network, and d) triangles in the network which are not $\delta$-instances of $M$, either due to edge order or time between first and last edge.}\label{fig:motifexample}
	\vspace{-0.5cm}
\end{figure}

Networks are ubiquitous models for real world systems, with applications ranging from social interactions to protein relationships \cite{newman2003structure}. 
%Many such systems are not static, but have an evolving structure over time \cite{holme2012temporal}. 
Many such systems are not static, but the edges are active only at certain points in time. 
The networks in which \emph{temporal edges} appear and disappear over time are called time-varying or \emph{temporal} networks. 
Examples of temporal networks include communication and transaction networks where each link is relatively short or instantaneous, such as phone calls or financial exchanges. 
%Another example is networks of physical proximity, capturing times at which individuals encounter each other. 

Time dependent and temporal properties can be analyzed on time-varying networks. Extracting recurring and persistent patterns of interaction in temporal networks is of particular interest, as it provides higher order information about the network transformation and functionality \cite{benson2016higher}.
%For example, in financial networks anomalous temporal patterns has been identified as early signals of topological collapse and financial crisis \cite{squartini2013early}.
For example, an abundance of triangles in financial transaction networks is associated with anomalies and is identified as the signature of financial crisis \cite{squartini2013early}. Similarly, phone calls have been shown to significantly increase in volume and change patterns during major events such as earthquakes \cite{yu2015analysis}.

Repeated patterns of interconnections between nodes occurring at a significantly higher frequency than those in randomized networks are called \textit{motifs}.
Formally, a \textit{temporal motif} is a subgraph and an ordering on the temporal occurrences of its edges. $\delta$-instances of a temporal motif are instances in which all the temporal edges appear according to the ordering specified by the temporal motif within a time window of length $\delta$ \cite{paranjape2017motifs,liu2018sampling}.
Figure \ref{fig:motifexample} illustrates a small temporal network with $\delta$-instances of a temporal motif $\bm$.

%While identifying significant temporal motifs is crucial for discovering network dynamics, it requires generating and counting the number of temporal motifs in ensembles of thousands of random networks. Therefore, it does not scale to modern temporal networks with tens of millions of edges.

However, currently no statistical models for temporal graphs exist that would allow for modeling frequencies of temporal motifs over time. Modeling of temporal motifs provides two challenges: First, one has to develop a statistical model of a temporal graph, which models temporal dynamics of node activations and who they send their edge/transaction to. And second, even after the model is built, the temporal motif frequencies and their variances need to be established. A naive approach would be to fit the model, generate/materialize a number of synthetic graphs from the model, and then apply expensive temporal motif counting algorithms in order to establish expected motif frequencies and their variances. Obviously such approach is infeasible and computationally too expensive and thus a new approach is needed.

Here we propose the{\em Temporal Activity State Block Model (\alg)}, which allows for efficient computation of expected temporal motif frequencies and their variances. \alg solves both above challenges: we provide an efficient model parameter fitting technique that scales {\em linearly} with the number of temporal edges in the graph; moreover, we also provide a {\em constant time} method for computing motif frequencies and their variances. This is in sharp contrast to naive approaches that would require exponential time.

The key to our approach is to develop a model where motif frequencies can be analytically derived and this computed in constant time (where the constant depends on the motif size and number of groups in \alg, both of which are fixed constants and usually small). 
\alg~ models different activity levels of groups of nodes in a temporal network and can effectively capture intermittent activation between individual nodes.
More precisely, our proposed \alg~model first partitions the nodes into different groups based on their activity level, \ie, the rate of temporal edges they are likely to send or receive.
We then model rate of out-links and in-links between every pair of groups using Poisson processes. Every pair of groups has distinct rate for sending and receiving temporal edges. The nodes' activity level, and hence their group assignment, may change over time.
%In real temporal networks, streams of edges often arrive in ``bursts'', resulting in sharp rises in the nodes' activity levels. 
Furthermore, \alg~ allows the rates of the Poisson processes to vary over time, and hence is able to robustly and efficiently model the bursty arrival of temporal edges that is observed in a real temporal network. Therefore, it allows for accurate identification of temporal network properties.

We conduct experiments on both synthetic and real-world temporal networks. 
%In our experiments, we compared the expected frequencies provided by our analytical framework to the actual motif counts for 2-node and 3-node motifs with 3 edges. 
%We first assess the accuracy of motif predictions with our analytical framework on synthetic networks generated according to the model. 
%We investigate the effect of varying the number of groups in TASBM, average degree of the temporal network, length of the time window $T$, and motif window $\delta$ on the accuracy of calculating the expected motif frequencies. 
Results on synthetic networks demonstrate that our analytical framework can closely track the changes in the frequencies of $\delta$-instances of temporal motifs over time.
When applying our framework to analyze synthetic networks containing planted anomalous motifs, our results show that although the planted motif anomalies cannot be identified based on motif counts alone, they can be discovered through comparison to the expected frequencies provided by our analytical model.

In our real-world experiments, we apply our analytical model to a subset of a financial transaction network with 118.7 thousand nodes and 2.9 million temporal edges.
In addition, we apply our framework to find trends in a phone call network with 1.2 million nodes and 21.9 million temporal edges, and a subset of an email network with 997 nodes and 307,869 edges. 
Our method can
accurately track motif frequencies in the most significant community of the financial network. In the full financial network, it successfully localizes the anomalies caused by a financial crisis. %We also observe that certain motifs can be interpreted as signals of improvement of the economy when the network starts to recover.
In the phone call network, we observe that  our analytical framework can identify weekly phone call trends such as peak hours and weekends.
%, and a widely observed holiday. %In addition, we can identify the temporal motifs that are important for the functionality of the network
%Furthermore, we can identify the patterns of business and family phone calls from the under-represented over the weekends,  are over-abundant during Good Friday, and .
%Furthermore, we can identify the patterns of business phone calls from under-abundance of certain motifs during the weekends, and patterns of personal phone calls from over-abundance of certain motifs on Good Friday.

\hide{
\xhdr{Contributions}
Here, we propose a temporal network model and use it to develop an analytic  model of temporal network motifs which does not require simulation. 
In particular:
\begin{description}
	\item[$\bullet$] We propose a temporal stochastic block model (TASBM) that captures different activity levels of groups of nodes in the network.
	\item[$\bullet$] We develop an efficient TASBM parameter fitting technique. 
	\item[$\bullet$] We derive closed-form expressions for calculating the expected number and the variance of $\delta$-instances of a temporal subgraph in large temporal networks.
    \item[$\bullet$] We apply our framework large real-world financial transaction and a phone call networks to discover trends and anomalies. We show that our model can discover trends and anomalies that are not identifiable by simply counting the number of edges, nodes, or motifs in the network.
\end{description}}

\section{Related Work}
\label{sec:related}
% !TEX root = motifs_newarxiv.tex

In this section, we review the related work to dynamic network and activity state models, as well as temporal motifs. 

\xhdr{Dynamic network models} 
There is a body of work on modeling dynamic networks, mostly by extending a static model to the dynamic setting~\cite{ahmed2009recovering,westveld2011mixed,ho2011evolving,yang2011detecting,decelle2011asymptotic,xu2013dynamic}. 
%Among the examples are temporal extensions of the exponential random graph model \cite{ahmed2009recovering} and latent space model \cite{westveld2011mixed}
%Two such models include temporal extensions of the exponential random graph model \cite{ahmed2009recovering} and latent space model \cite{westveld2011mixed}. 
Among the existing methods, temporal extensions of stochastic block models (SBMs) are the most relevant to our work. In particular, 
%More closely related to the state-space model we propose are several temporal extensions of stochastic blockmodels (SBMs).
a dynamic SBM, in which a transition matrix specifies the probability of nodes to switch classes over time \cite{yang2011detecting,ho2011evolving}.
%involving a transition matrix that specifies the probability that a node in class $i$ at time t switches to class $j$ at time $t+1$ for all $i, j, t$ and fit the model using Gibbs sampling and simulated annealing. 
%Ho et al. proposed a temporal extension of a mixed-membership SBM (MMSBM) using linear state-space models for the class membership vectors of node clusters.
However, in contrast to our work here the above works both assume that edge probabilities do not change over time. 
%Both \cite{yang2011detecting, ho2011evolving} treat edge probabilities as time-invariant parameters.
More recently, models in which the network snapshots are modeled using SBM \cite{xu2013dynamic,decelle2011asymptotic}, and the state evolution of the dynamic network is modeled by a stochastic dynamic system have been proposed.  
Other temporal extensions of block model include \cite{corneli2015modelling}, which presents a non-stationary extension of the SBM; and \cite{matias2017statistical} that performs clustering on temporal networks, and uses Markov chains to model node groups.
%. Matias et al.~\cite{matias2017statistical} perform clustering on temporal networks, use Markov chains to model node groups in an extension of the SBM.
Our work here differs in that it focuses on temporal graphs, the model allows for linear time fitting and provides closed form expressions for temporal motif counts.

%Our work here differs from the above in that in our proposed \alg~ model, a particular block is not a community with vertices more likely to be connected to each other, but a set of vertices with similar local behavior, in terms of out- and in- edge arrival rates. 

\xhdr{Temporal motifs} 
Paranjape et. al.~\cite{paranjape2017motifs} extended the concept of static motifs to temporal networks and proposed a framework for counting the exact number of relatively small temporal motifs. Very recently, \cite{liu2018sampling} proposed sampling methods for approximating the number of temporal motifs.
Existing methods either do not account for ordering of the temporal edges \cite{zhao2010communication},  or require temporal edges in a motif to arrive consecutively to a node \cite{kovanen2011temporal}. Among other examples are the algorithm of \cite{gurukar2015commit} that uses ideas from sub-sequence mining to identify patterns in temporal graphs, and methods of \cite{redmond2013temporal} for finding temporal isomorphic subgraphs. 
While this line of work aims to accurately count or approximate the actual numbers of motif instances in any graph, the goal of our work is to determine the expected number and variance of motif instances in a graph, given its underlying statistical model. 

%\textbf{Null models for temporal networks.}
Currently no statistical model of motifs in temporal graphs exists. However, there have been several heuristic/empirical techniques proposed.
Such approaches are based on generating a large ensemble of randomized temporal networks and include shuffling and reversing of timestamps~\cite{bajardi2011dynamical,donker2014dispersal,holme2014birth,kovanen2011temporal,li2018lifetime}.
%
%Bajardi et al.~\cite{bajardi2011dynamical} showed the inherent ordering of motifs due to their disappearance in shuffled or reverse-ordered time series. Donker et al.~\cite{donker2014dispersal} used time reversal in their analysis of a hospital network. Kovanen et al.~\cite{kovanen2011temporal} compared temporal network data to randomly permuted timestamps as well as permuted timestamps with bias toward shorter inter-event times.  Holme et al.~\cite{holme2014birth} and Li et al. ~\cite{li2018lifetime} presented models of temporal randomization with additional constraints to preserve the lifetimes of edges. 
%
Unlike these heuristic/empirical methods, we develop a statistical model and closed-form expressions for motif counts and their variances. Crucially, our approach does not require expensive shuffling/simulation of network ensembles, and this scales well to large temporal networks.
%\jure{Again, explain how our work relates to this line of work}

% !TEX root =motifs_newarxiv.tex

\newcommand{\blockparams}{$(C^{in},C^{out},\mathcal{G}^{in},\mathcal{G}^{out},\mathcal{R})$}

\section{Network Activity Model}
In this section we describe our Temporal Activity State Block Model (\alg). Our goal is to construct a temporal network model which can be updated efficiently and thus maintained online, so that it can be used to describe the network before the full edge set is known.

Formally, a \emph{temporal graph} can be viewed as a sequence of static directed graphs over the same (static) set $V$ of $n=|V|$ nodes and the set $E$ of $m=|E|$ \emph{temporal edges}. 
Each \emph{temporal edge} is a timestamped ordered pair of nodes $(e_i=(u, v), t_i),  i \in [m]$, where $u, v \in V$ and $t_i \in \mathbb{R}$ is the timestamp at which the edge arrives. For example, in a phone call network, each temporal edge includes the caller, $u$, the receiver, $v$, and the time, $t_i$ at which the call was placed.
Multiple temporal edges between the same pair of nodes $u$ and $v$ and different timestamps can exist. We assume that the timestamps $t_i$ are unique so that the temporal edges may be strictly ordered.  However, our methods do not rely on this assumption and can easily be adapted to the case where timestamps are not unique, e.g. by consider each possible ordering of edges with a shared timestamp or selecting an order at random.

The \emph{stochastic block model} \cite{decelle2011asymptotic} on static networks is defined as dividing the nodes into communities, or blocks, such that a higher proportion of the possible edges within a block occur, compared to those between blocks. %The model then represents the graphs as a set of blocks with edge probabilities within and between each block. Taking this approach on temporal networks is not realistic because communities evolve over time and determining them at frequent time intervals is prohibitively expensive. 
We propose a temporal variant of the stochastic block model in which we
partition the nodes of the network into groups based on their activity levels, which we define as the rates at which in- and out- edges arrive to nodes. This means that a particular block is %not a \emph{community} with vertices more likely to be connected to each other, but 
a set of vertices with \emph{similar temporal activity}. Specifically, nodes within each block will all have similar rates of out- and in- edge arrivals.

\subsection{Temporal Activity State Block Model}
In our Temporal Activity State Block Model (\alg) we consider two sets of groups or activity states $G^{in}=\{1,\cdots, C^{in}\}, G^{out}=\{1,\cdots, C^{out}\}$. %, where $C^{in} = |G^{in}|$ and $C^{out}=|G^{out}|$. 
Every node $u$ in the network belongs to a group $a^{in}_u = i \in G^{in}$  based on its activity level for receiving in-links and a group $a^{out}_u = j \in G^{out}$ based on its activity level for sending out-links. 
Nodes in the same group $i \in G^{in}$ have similar rate of receiving temporal edges. Similarly, nodes in the same group $j \in G^{out}$ have similar rate of sending temporal edges. %%%We denote by $\theta_i$ the expected rate of in-links or out-links for group $i$.
We model group assignments $a^{in}_u, a^{out}_u$ for $u \in V$ as independent draws from multinomial distributions parameterized by $\pi^{in}, \pi^{out}$.  %so $\alpha_r = P(g_u = r)$ is the prior probability that a node is in block $r$. 
Thus, $a^{in}_u \sim \text{Multinomial}(\pi^{in})$, and $a^{out}_u \sim \text{Multinomial}(\pi^{out}).$
%%

%%%We then consider a $C^{in} \times C^{in}$ matrix $\Theta^{in}$, where $\theta^{in}_{ij}$ denotes the rate of temporal edges group $i$ received from group $j$, and another $C^{out} \times C^{out}$ matrix $\Theta^{out}$, where $\theta^{in}_{ij}$ denotes the rate of temporal edges group $i$ sends to group $j$. 

%There are $C = C^{out} \times C^{in}$ states in our activity-state block model, and at every time $t$ each node $u$ belongs to one of the states $s_u \in \{1, \cdots, C\}$.

We consider a $C^{out} \times C^{in}$ matrix $\pmb{\theta}$ such that $\theta_{ij}$ denotes the rate of temporal edges from nodes in group $i \in G^{out}$ to the nodes in group $j \in G^{in}$.
After assigning nodes to different activity states, we  model  the temporal edges between every pair $(u,v)$ of nodes with $a^{out}_u = i, a^{in}_v = j$ as independent Poisson draws, where the means of these Poisson draws are specified by $\theta_{ij}$. More formally, every temporal edge $(e_r={(u,v)}, t_r)$ between the node $u$ in out-link activity state $a_u^{out}=i$ to the node $v$ in in-link activity state $a_v^{in}=j$ is an independent Poisson draw. I.e., 
$$e_r={(u,v)}|a^{in}_u\!=\!i, a^{out}_v\!=\!j \sim \text{Poisson}(\theta_{ij}).$$
For the ease of notation, instead of $\theta_{a_u^{out} a_v^{in}}$, we subsequently use $\theta_{a_u^{} a_v^{}}$ to denote the rate of out-links from nodes in activity state $a_u$ to the nodes in activity state $a_v$. %\textbf{TODO: Note that the choice of Poisson draws is not required: any method of generating edges with some expected value $\theta_{a_u,a_v}$ can be used in the analysis that follows.}

%It has been observed that in real temporal networks, stream of  edges usually arrive in bursts, resulting in sharp rises in the nodes' activity levels \cite{kleinberg2003bursty,holme2012temporal}. To capture this activity, 

As nodes change their activity level over time, we assign individual nodes to a state based on their activity levels on each time window $T$.
Then we model arrivals of temporal edges between each pair of nodes as a Poisson process with a constant parameter on time intervals of length $T$. %, i.e. $\Theta$ is a constants matrix in $[t_0, t_0+T)$. 
Note that the rates of the Poisson processes can vary for each time window, and hence \alg~ is able to
%The \alg~ allows us to
robustly and efficiently model the bursty arrival of temporal edges that is observed in a real temporal network.

\subsection{Parameter Inference in \alg}
For all pairs of vertices $(u,v)$ such that $a^{out}_u = i$ and $a^{in}_v = j$ for some $i \in G^{out}$ and $j \in G^{in}$, the Poisson process modeling temporal edges between $u$ and $v$ will be parameterized by a constant $\theta_{ij}=\theta_{a_ua_v}$ in every time interval of length $T$. Across intervals, we calculate the posterior model parameters $\alpha$ and $\theta$ as:
$$\hat{\alpha}=\frac{n_r}{n}, \hat{\theta}_{rs}=\frac{m_{rs}}{n_r n_s},$$
where we denote by $n_r$ the number of nodes in group $r$, and by $m_{rs}$ the number of edges connecting group $r$ to group $s$ in a time interval of length $T$.
Model inference can be done in at most two passes over the edges and in practice, can be well-approximated in one pass (see Section~\ref{sec:fit} for details). Thus, this method is extremely scalable and runs in time {\em linear} in the number of edges in the graph.
%Regardless of its simplicity we find the method performs well in practice and scales to large datasets.

% !TEX root = motifs_newarxiv.tex
\section{Analytical Model for Temporal Motifs}\label{sec:allmodelanalysis}

%While we can easily infer the parameters of a \alg, identifying significant motifs requires generating random ensembles of thousands of networks with the inferred parameters and counting motifs in each network. This is impractical for large temporal networks, and we need to resort to analytical solutions.

Having defined the model our next task is to ``count'' the motifs. In fact, we do not want to count them as counting would mean materializing/sampling many networks from the model and then running expensive motif counting algorithms. Rather, we will analytically derive closed-form expressions that allow us to quickly calculate motif counts and their frequencies.

We first provide formal definitions of temporal motifs and $\delta$-instances of temporal motifs. We then introduce our analytical framework for calculating expectation and variance of the number of motif instances. %Such a framework is necessary because it is not computationally feasible to sample a network in the \alg and count the motifs. 
Finally, we provide the computational complexity of our method and show that it can easily scale to large real-world temporal networks with millions of temporal edges.

\subsection{Temporal Network Motifs}\label{sec:pre}
Formally, a temporal motif $\bm$ defines a particular sequence of interactions between a set of nodes over time. 
\begin{definition}[temporal motif]
A $k$-node $z$-edge temporal motif $\bm = (G_{\bm},\prec_{\bm})$ consists of a graph $G_{\bm}=(V_{\bm},E_{\bm})$, such that $|V_{\bm}|\!=\! k$ and $|E_{\bm}|\!=\!z$, and a strict total ordering $\prec_{\bm}$ on the edges $E_{\bm}$. We index $E_\bm = \{e'_1, \!\allowbreak \cdots\!, e_z' \}$, such that $e'_1 \prec_{\bm} e'_2 \prec_{\bm} \!\cdots \!\prec_{\bm} e'_z$. 
\end{definition}

Note that multiple interactions between the same pair of nodes may occur in the sequence defined by $\bm$, but each edge is indexed and ordered uniquely. In a dynamic network, any subgraph of a temporal graph is a $\delta$-instance of a temporal motif $\bm$ if it is isomorphic to $G_\bm$, the set of its temporal edges follows the same ordering imposed by $\bm$, and it occurs within a time window of $\delta$. 

\begin{definition}[$\delta$-Instance of a temporal motif]\label{def:instance}
	A temporal subgraph $G_s=(V_s, E_s)$, with $E_s=\{(e_1, t_1), \cdots, (e_z, t_z)\}$ 
	is a $\delta$-instance of a temporal motif $\bm$ if 1) \textit{isomorphism}: there exist an edge-preserving bijection  $f: V_s  \rightarrow V_\bm$ between nodes of the subgraph and nodes of the motif such that $\forall_{e=(u,v) \in E_s }(f(u),f(v)) \in E_\bm$; 2) \textit{temporal ordering}: the edges of the temporal motif occur according to the ordering $\prec_{\bm}$, i.e.,
	for the ordered sequence $f(e_h) \prec_{\bm} f(e_i) \prec_{\bm} \cdots \prec_{\bm} f(e_j)$ we get a set of strictly increasing timestamps $t_h < t_i < \cdots < t_j$; and 3) \textit{temporal window}: all the edges in $E_s$ occur within $\delta$ time, i.e. $t_j - t_h \leq \delta$.
\end{definition}

Here, our goal is to derive the expected number (and the variance) of $\delta$-instances of a given temporal motif in a time-varying network. More precisely, given the number of nodes and degree distribution of a temporal network at each point in time, we are interested in calculating the expected frequency of ordered subsets of edges from the temporal network that are $\delta$-instances of a particular temporal motif. 
In the following, we present a general analytic approach, 
%to replace timestamp shuffling and rewiring, 
and show how to apply this approach to the Temporal Activity State Block Model (\alg).

 %\subsection{Proposed Method} \label{sec:proposed}

%\vspace{.1cm}\textbf{Considerations.} 
%We note that while the number of nodes and pairs of connected nodes can be of manageable size, the number of temporal edges may be very large and thus efficient algorithms are needed when analyzing such data.
%
%In small temporal networks, the expected temporal motif frequencies can be calculated by 
%Existing approaches for calculating the expected temporal motif frequencies work by 
%re-wiring and shuffling the sequence of temporal edges, and actually counting the number of instances of motifs in the resulting randomized networks \cite{bajardi2011dynamical}. % Similarly, non-analytical methods for calculating expected static motifs work by re-wiring edges while preserving the degree of vertices, and counting motifs in these randomized networks.
%However, in real temporal networks the number of edges is quite large, and counting the number of motifs on the set of shuffled and/or re-wired networks is prohibitive. 

\subsection{Expected Motif Frequencies for \pmb{$T \leq \delta$}}\label{sec:analytical}
%In this section, we present our analytical framework for calculating the expected number of $\delta$-instances of temporal motifs in a temporal network. %We first consider the most general setting, with time-varying edge arrival rate defined for each pair of nodes. 
Next we provide a closed form solution for the expected $\delta$-instances of temporal motifs in \alg~ networks. 
%We note that, our analysis is not limited to \alg~ and can be easily generalized to other temporal network models. In the most general case, every node has a separate activity state, and hence we have a \alg~ with $n^2$ activity states. %In the most general case, every node %\textbf{TODO: see appendix for generalized analysis} %The notation used in this section is summarized in Table \ref{table:notations}.
 %We then show how the expected number of instances can be computed using the parameters of the TASBM.
%
%We compute the expected number of $\delta$-instances of a temporal motif $M=(G_M,\prec_M)$ under an arbitrary network model.
%As discussed in Section \ref{sec:pre}, in a temporal graph multiple edges with different timestamps may appear between the same pair of nodes. 
%For every pair $(u,v)\in V\times V$, we define the edge arrival rate $\theta_{{uv}} \in \R$ as the number of temporal edges from u to v occurring per time unit.
%More precisely, for every time window $W = [t_0, t_0+T)$ of length $T$, we have $\theta_{{uv}} = \sum_{t_i \in W} \mathbb{I}[(e_i=(u,v), t_i)]/T$. 
%We begin by assuming that the temporal edges arrive with a constant rate between all pairs of nodes in the network, i.e., $\theta_{e_{i}} = \lambda$ for all $e_i \in V\times V$. We then consider the more general scenario where each pair of nodes has a distinct edge arrival rate. %which may be $0$.
%
To calculate the expected number of $\delta$-instances of a temporal motif over a time window of length $T$, we need to calculate the expected number of subgraphs $G_s=(V_s, E_s)$ satisfying the conditions specified in Definition \ref{def:instance}. %I.e., we need to 1) compute the number of possible isomorphic subgraphs to the motif graph $G_\bm$; 2) calculate the probability that timestamps of the edges in each isomorphic subgraph occur in the order $\prec_M$ specified by the motif and 3) all the edges of the subgraph occur within a time window of length $\delta$. 
The following Theorem summarizes our main theoretical results.
\begin{theorem}\label{thm1-1}
	The expected number of $\delta$-instances of a $k$-node $z$-edge temporal motif $\bm$ in a network with a set $V$ of nodes modeled by a \alg~ with $C$ states during a time interval $[t_0, t_0+T)$ for $T \leq \delta$ is
	$$ \E[N_\bm|{T\leq\delta}]= \E[N_{S_{V,C}^{k,z}}] \cdot Pr(t_0\leq t_1<t_2<...<t_z< t_0+T), $$
	where the first term is the expected number of $k$-node $z$-edge isomorphic subgraphs to the motif graph $G_\bm$ in a temporal network with $V$ nodes modeled by a \alg~ with $C$ states;
	and the second term is the probability that timestamps of the edges in each isomorphic subgraph occur in the order $\prec_M$ specified by the motif in a time window of length $\delta$.
\end{theorem}

In the rest of this section, we explain in detail how we derive the above quantities and calculate a closed form solution for the expected number of $\delta$-instances of a temporal motif.

\xhdr{Expected Number of Isomorphic Subgraphs} %\label{sec:isomorphicsub}
We start by calculating the number of ways that subgraphs isomorphic to the motif graph $G_\bm$ may occur in the temporal network.
In a static network, the number of subgraphs isomorphic to $G_\bm$ is equal to the number of unique edge-preserving bijections between nodes of the network and nodes of the motif $\bm$. However, in a temporal graph multiple edges may appear between every pair of nodes in a time interval of length $T$, and therefore each bijection may result in multiple isomorphic subgraphs to $G_\bm$.
To calculate the number of isomorphic subgraphs, we first compute the number of unique  bijections %$f:V_S\to V_M$ between all $k$-node $z$-edge subgraphs $G_s=(V_s,E_s)$ and $G_\bm$. 
$f: V_s  \rightarrow V_\bm$ between the nodes of every $k$-node $z$-edge subgraphs $G_s=(V_s,E_s)$ and $G_\bm=(V_\bm,E_\bm)$. %such that $\forall e=(u,v) \in E_s \Leftrightarrow f(e=(u,v)) \in E_\bm$.
Then we compute the number of isomorphic subgraphs that may result from each bijection $f$ in a time interval of length $T$.

%Without partitioning the nodes to different activity states, 
In general, in a graph with $n$ nodes, there are ${n \choose k}$ ways to choose nodes to form a $k$-node subgraph. Furthermore, there are $k!$ permutations of a set of $k$ nodes, each corresponding to a unique bijection from $V_\bm$ to $V_s$. For example, the motif in Figure \ref{fig:1b} has $3!=6$ possible vertex bijections to a $3$-node $3$-edge graph (see Appendix Figure~\ref{fig:permutationex}).

  %removes any such redundancy.
% To count the number of edge-preserving bijections between nodes of the subgraph $V_s$ and nodes of the motif graph $V_\bm$, we preserve the pattern of connectivity defined by the motif $\bm$, i.e., we maintain the adjacency matrix of the static subgraph induced by $E_\bm$, and permute the nodes of the subgraph $V_s$. In a directed subgraph, for each permutation of the $k$ nodes in $V_s$ we get a bijection.
Therefore, the number of bijections $\mathcal{B}_{k,V}$
%$$\mathcal{B}_{n,k} = \{f: V_s  \rightarrow V_\bm | k=|V_s|, \forall e=(u,v) \in E_s \Leftrightarrow f(e=(u,v)) \in E_\bm\}$$
between nodes of the temporal network and nodes of the motif is %${n \choose k} k! = P(n,k)$. We let the set $B_{k}$ denote the set of such bijections, so 
$$|\mathcal{B}_{k,V}| = {n \choose k} k! = P(n,k).$$

%%%%%%%%%%%%%%%%%%%%%%%%
We now derive the number of ways that subgraphs isomorphic to the motif graph $G_\bm$ may occur in the Temporal Activity State Block Model (\alg).
Here, we first calculate the number of ways nodes in different activity states can form a $k$-node subgraph.
%of a $k$-node subgraph can be selected from $C$ activity states.
%for each of the $k$ nodes in the subgraph, we first select an activity state, and then select a node in that activity state.
Any node in the subgraph can be selected from at most $C$ activity states. 
%Figure \ref{fig:assignment} shows two different activity state assignment to the nodes of a 3-node subgraph.
Hence, the number of activity state assignments $\mathcal{A}_{C,k}$ for $k$ nodes in \alg~ is\footnote{If some groups have fewer than $k$ nodes, we have $\mathcal{A}_{C,k}<C^k$. For example, let $k = 3$ and $s_i$ be the number groups of size $i$ for $i\in\{0,1,2\}$, then $\mathcal{A}_{C,k} = (C-s_0)^k -s_1{k\choose 2}(C-1-s_0)-s_2$. If each node is in a separate group with $C=|V|$ we get $|\mathcal{B}_{k,V,n}| = |\mathcal{B}_{k,V}| = P(n,k).$} %from nodes in $C$ activity states is 
% are at most $C^k$ ways to select $k$ nodes from the $C$ activity states in the subgraph
$$ |\mathcal{A}_{C,k}^{}| \leq C^k$$

To compute the number of bijections, 
assume that nodes of the temporal graph are partitioned into $C$ activity states as $ V\!=\!\{V^1, \!\cdots, \!V^C \} $, and let $n^c=|V^c|$ be the number of nodes in activity state $c \in [C]$. %, i.e., $n^c = |V^c|$. %, where $a_v \in [C]$ is the activity state of node $v$.
Consider an activity state assignment $A=\{a_1, \cdots, a_k\} \in \mathcal{A}_{C,k}$, where $a_i \in [C]$ is the activity state of the $i$-th node. 
We now calculate the number of subsets $V_s$ of $k$ nodes in the network that are consistent with the activity state assignment $A$ (see Appendix Figure~\ref{fig:isomorphic} for an example of activity states mapped to nodes).
Let $n^c_{A}$ be the number of nodes in $A$ that are assigned to the activity state $c$. %, i.e. $n^c_{A} = \sum_{i\in[k]} \{a_i =c |a_{i} \in A\}$. 
There are ${n^c \choose n^c_{A}}$ ways to select $n^c_{A}$ nodes from $V^c$. Therefore, there are $ {n^1 \choose n^1_{A}} \cdots {n^C \choose n^C_{A}}  $ ways of forming a $k$-node subgraph in the network that are consistent with the activity state assignment $A$. %such that $G(V_s)=A$. 
%Figures \ref{fig:assignment} shows an example. 
%
Finally, for each permutation of the $n^c_{A}$ nodes in activity state $c$ we get a bijection.
Hence, the number of unique bijections in TASBM is 
\begin{equation}\label{eq:group_count}
%N^c_{S_{k,z}} = 
|\mathcal{B}_{k,V,C}^{}| = \sum_{\substack{A \in \mathcal{A}_{C,k}^{}}} \prod_{c \in [C]} P(n^c, n^c_{A}), %, \cdots P(n^C, n^C_S),
\end{equation}
where $P(n^c\!,\! n^c_{A})\!\! =\!\! {n^c \choose n^c_{A}} n^c_{A}!$ is the number of $n^c_{A}\!$-permutations of $n^c$. 
Note that $\prod_{c \in [C]} P(n^c, n^c_{A})$ is constant for every activity state assignment $A \in \mathcal{A}_{C,k}$, and hence the cost of calculating Eq. \ref{eq:group_count} only depends on the number of possible activity state assignments $O(C^k)$.

%In a static network, the number of isomorphic subgraphs to $G_\bm$ is equal to the number of unique bijections between nodes of the network and nodes of the motif $\bm$. However, in a temporal graph multiple edges may appear between every pair of nodes and therefore each bijection may result in multiple isomorphic subgraphs to $G_\bm$.

%For every possible edge $e_i=(u,v)$ from node $u$ to node $v$, we denote temporal edge arrival rate at time $t$ by $\theta_{e_i}(t)$. Hence, the expected number of temporal edges that occur by time $T$ from $u$ to $v$ is $\int_0^T \theta_{e_i}(t)dt$. 

Next, we calculate the expected number of isomorphic subgraphs that can result from each bijection in a time interval of length $T$.
As we discussed in Section \ref{sec:pre}, in the \alg~ the arrival rate of temporal edges between any pair of nodes $(u,v)$ depends on their activity states. 
%In particular the edge arrival rate from node $u$ in activity state $a_u$ to node $v$ activity state $j$ is $\theta_{a_ua_v}$.
The expected number of temporal edges that occur from node $u$ in activity state $a_u$ to node $v$ in activity state $a_v$ in an interval $[t_0, t_0+T)$ is 
$$ \E[N_{e_{uv}}|t\in[t_0,t_0+T)] = \int_{t_0}^{t_0+T} \theta_{{a_ua_v}}(t)dt,$$ where $ \theta_{{a_ua_v}}(t)$ is the temporal edge arrival rate from activity state $a_u$ to activity $a_v$ at time $t$.

%For every activity state assignment $A \in \mathcal{A}_{C,k}, A(u) \in [C]$, %$a = \{ a_1, \cdots, a_k \} \in \mathcal{A}_{C,k}$, all the  $\prod_{c \in [C]} P(n^c, n^c_{V_s})$  
%all the bijections corresponding to $a$ have similar edge arrival rates between their nodes and hence have equal  the expected number of subgraphs are equal.

Let $f: V_s \rightarrow V_\bm$ be a bijection, where $V_s$ is a $k$-node subgraph that is consistent with the activity state assignment $A\in \mathcal{A}_{C,k}$. In other words, $V_s=\{v_1, \cdots, v_k |a_{v_i}=A[i] \}$. %corresponding to an activity state $A=\{a_{v} | v \in V_s\} \in \mathcal{A}_{C,k}$.
%Let $A \in \mathcal{A}_{C,k}$ be an activity state assignment for a $k$-node subgraph and $V_a = \{ v_1, \cdots, v_k \}$ be a subgraph in which $v_i \in V^{c_i}$.
%Moreover, assume there is an edge-preserving bijection $f: V_a \rightarrow V_\bm$.
We now calculate the expected number of $k$-node $z$-edge isomorphic subgraphs $S^{k,z}_{V,C,f}$ that can result from $f$ in a \alg~ with nodes $V$ partitioned into $C$ activity states, in a time interval of $[t_0, t_0+T)$.
%For every group assignment $a$ the expected number of subgraphs that are isomorphic to the static subgraph $G_\bm$ specified by the motif, in a time interval $[t_0, t_0+T)$ is
\begin{equation}\label{eq:dyn_num}
%\E[N_{S_{k,z}}] = \sum_{\substack{f \in B_k, \\f: V_s \rightarrow V_\bm}} 
\E[N_{S^{k,z}_{V,C,f}}|t\in[t_0,t_0+T)] = 
\!\!\!\!\prod_{\substack{u,v \in V_s,\\(f(u),f(v)) \in E_\bm}} \int_{t_0}^{t_0+T} \theta_{a_ua_v}(t)dt.
\end{equation}

Note that for each activity state assignment $A \in \mathcal{A}_{C,k}$, 
the expected number of subgraphs for all the $\prod_{c \in [C]} P(n^c, n^c_{s})$ bijections corresponding to $A$ is the same.
%For every edge-preserving bijection from the set $V_s$ of $k$ nodes in the network to the set of nodes of the motif $V_\bm$, i.e., $f: V_s \rightarrow V_\bm$ such that $f(e_i) \in E_\bm$, 
%Let $S_{k,z}$ be the set of all $k$-node $z$-edge subgraphs isomorphic to $G_\bm$ in the temporal network, each corresponding to a bijection in $B_k$. 
%where by overloading notation, we use $\theta_{e_i=(u,v)}=\theta_{g_u^{out}, g_v^{in}}$.
Therefore, from Eq. \ref{eq:group_count} and \ref{eq:dyn_num} we get the following lemma.

\begin{lemma}\label{lemma1}
	The expected number of $k$-node subgraphs isomorphic to the motif subgraph $G_\bm$  in a Temporal Activity State Block Model (TASBM) during a time interval $[t_0, t_0+T)$ for $T \leq \delta$ is
	\begin{align*}%\label{eq:dyn_num-}
	%\E[N_{S_{k,z}}] = \sum_{\substack{f \in B_k^{BM}, \\f: V_s \rightarrow V_\bm}} \prod_{\substack{e_i \in E_s,\\f(e_i) \in E_\bm}} \int_0^T \theta_{e_i}(t)dt .
	%\E[N_{S_{k,z}}] = \sum_{\substack{A \in \mathcal{A}_{C,k}}} F(g) \prod_{\substack{(u,v) \in E_s,\\f(u,v) \in E_\bm}} \int_0^T \theta_{A(u),A(v)}(t)dt .
	\E[N_{S^{k,z}_{V,C}}&|t\in{[t_0,t_0+\delta)}] = \\
	&\sum_{\substack{A \in \mathcal{A}_{C,k}^{} } } %,\\  a_v \in [C] } }% a:V_s \rightarrow \{a_v \in [C] | v \in V_s \} } }%a=\{ a_v\in [C] | v\in V_s \} } } 
	 \prod_{c \in [C]} P(n^c, n^c_{s}) 
	\!\!\! \prod_{\substack{u,v \in V_s|R(V_s)=A,\\ (f(u),f(v)) \in E_\bm}} \int_{t_0}^{t_0+T} \theta_{a_ua_v}(t)dt,
	\end{align*}
	where $R(V_s)=A$ is the set of all $k$-node subgraphs consistent with activity state assignment $A$. 
\end{lemma}

\textbf{Probability of Temporal Ordering for $\pmb{T \leq \delta}$.} 
We now compute the probability that the second and third conditions in Definition \ref{def:instance} are satisfied. Precisely, we calculate the probability that for an isomorphic subgraph $G_s =(V_s, E_s)$, the timestamps of the mapped edges ordered by $\prec_M$ are strictly increasing and within a time window of $\delta$. 

For ease of notation, we subsequently assume that for each subgraph $G_S$ and corresponding bijection $f:V_s \to V_{\bm}$, we have $f(e_1) \prec_M f(e_2) \prec_M \cdots \prec_M f(e_z)$. Therefore, we need to calculate the probability that $t_0\leq  t_1 < t_2 < \cdots < t_z < t_0+\delta$.

The marginal probability for a temporal edge $e=(u,v)$ from activity state $a_u$ to activity state $a_v$ to occur at time $t$ in the time window $[t_0,t_0+T)$ is 
$$\Theta^{[t_0,t_0+T)}_{{e=(u,v)}}(t) =\frac{\theta_{{a_ua_v}}(t)}{ \int_{t_0}^{t_0+T} \theta_{{a_ua_v}}(t')dt'}.$$

%We define the probability distribution each edge from group $i$ to group $j$ on interval $[t_1,t_2]$ as $$\Pi_{ij}^{[t_1,t_2]}:= \frac{\pi_{ij}(t)}{ \int_{t_1}^{t_2} \pi_{ij}(t)dt}.$$

\begin{lemma}\label{lemma2}
The probability that temporal edges of a subgraph $G(V_s, E_s)$ occur in the order $\prec_\bm$ specified by motif $\bm$ in an interval $[t_0, t_0+T)$ is
\begin{align*}%\label{eq:pr_dyn}
\hspace{-1.2cm}Pr(t_0\leq t_1<t_2<...&<t_z< t_0+T) = \\
\int _{t_0}^{t_0+T} &\Theta^{[t_0,t_0+T)}_{{e_1}}(t_1)  \int_{t_1}^{t_0+T} \Theta^{[t_0,t_0+T)}_{{e_2}}(t_2) \cdots \nonumber\\
&\int_{t_{z-1}}^{t_0+T} \Theta^{[t_0,t_0+T)}_{{e_z}}(t_z) dt_{z} dt_{z-1} dt_{z-2}...dt_1	\nonumber.
\end{align*}

\end{lemma}
%\textbf{[Proof in Appendix.]}
Note that in the above equation, if the edge arrival rates $\theta_{ij}$ between activity states do not change in interval $[t_0, t_0+T)$, the marginal probability for every temporal edge to happen is $1/T$. Hence, every ordering of the temporal edges in the subgraph has a probability of $1/z!$. Therefore, for constant edge arrival rates we have  $Pr(t_0\leq t_1<t_2<\cdots<t_z< t_0+T) = 1/z!$. On the other hand, varying edge arrival rates in $[t_0, t_0+T)$ can be modeled by integrable functions, for which we can calculate the value of the nested integrals to get the probability of the correct ordering $Pr(t_0\leq t_1<t_2<\cdots<t_z< t_0+T)$.

Finally, the expected number of $\delta$-instances of a motif $\bm$ is the number of subgraphs that satisfy all the conditions specified in Definition \ref{def:instance}. It can be calculated by multiplying the number of possible isomorphic subgraphs to the motif graph $G_\bm$ by the probability that timestamps of the edges in each isomorphic subgraph occur in the order $\prec_M$ specified by $\bm$ within a time window of length $\delta$.

From Lemma \ref{lemma1} and Lemma \ref{lemma2} we get %the following theorem
Theorem \ref{thm1-1} for the expected number of $\delta$-instances of temporal motifs %$\bm$ 
in a time interval of length $T\leq\delta$.
%\begin{theorem}\label{thm1}
%	The expected number of $\delta$ instances of a temporal motif $\bm$ in a Temporal Activity State Block Model (\alg) during a time interval $[t_0, t_0+T)$ for $T \leq \delta$ is
%	$$ \E[N_\bm|{T\leq\delta}]= \E[N_{S_{V,C}^{k,z}}] \cdot Pr(t_0\leq t_1<t_2<...<t_z< t_0+T). $$
%	%\begin{align}
%	%\E[N_\bm|{T\leq\delta}]= E[N_S]
%	%\sum_{\substack{g \in G_{n,k,C}^{}}} F(g) \prod_{\substack{e_i \in E_s,\\f(e_i) \in E_\bm}} \int_{t_0}^{t_0+T} \theta_{e_i}(t)dt \nonumber\\
%	%\hspace{1cm} \cdot Pr(t_0\leq t_1<t_2<...<t_z< t_0+T)]. \nonumber
%%	&\int _{t_0}^{t_0+T} \Theta^{[t_0,t_0+T)}_{e_1}(t_1)  \int_{t_1}^T \Theta^{[t_0,t_0+T)}_{e_2}(t_2) \cdots \nonumber\\	
%%	&\hspace{1cm}\int_{t_{z-1}}^T \Theta^{[t_0,t_0+T)}_{e_z}(t_z) dt_{z} dt_{z-1} dt_{z-2}...dt_1	\nonumber.
%	%\end{align}
%\end{theorem}
%
The pseudocode of our method is given in Alg. \ref{alg:atsbm}.

%%We now use Eq. \ref{eq:dyn_num} and Eq. \ref{eq:pr_dyn} to calculate the expected frequency of $\delta$-instances of temporal motif $\bm$ with $T\leq \delta$:
%%%\begin{lemma} The expected frequency of $\delta$-instances of temporal motif $\bm$ in time window $T\leq \delta$ is
%%\begin{align}\label{eq:exp_dyn}
%%\E [N_\bm|T\!\leq\!\delta] = 
%%&\sum_{\substack{f \in B_k, \\f: V_s \rightarrow V_\bm}}\Bigg[\prod_{\substack{e_i \in E_s,\\f(e_i) \in E_\bm}} \left( \int_0^T \theta_{e_i}(t)dt \right) \! \nonumber\\
%%&\hspace{1.5cm} \cdot  Pr (t_0\! \leq \! t_1 \! < \! t_2 \!<\! \cdots \!<\! t_z \!< \! t_0\!+\!T) \Bigg]
%%%&\sum_{G_s \in I_\bm} \prod_{i =1}^z \int_0^T \theta_{e_i}(t)dt\\
%%%&\int _0^T \Theta^T_{e_1}(t)  \int_{t_1}^T \Theta^T_{e_2}(t_2)... \!\!\int_{t_{z-1}}^T \Theta^T_{e_z}(t_z) dt_{z} dt_{z-1} dt_{z-2}...dt_1	
%%\end{align}

\subsection{Expected Motif Frequencies for $\pmb{T > \delta}$}
Previously, we calculated the probability of the temporal ordering specified by $\bm$ for the case where $T \leq \delta$.
Next, we consider the scenario where $T > \delta$.

Here, the $\delta$-instances of a temporal motif may have at least one edge occurring in $[t_0,t_0\!+\!T\!-\delta)$ or they may have all edges occurring in $[t_0\!+\!T\!-\!\delta,t_0\!+\!T)$. Hence, to calculate the expected number of instances in $T$, we take a sum over the instances with at least one edge in $[t_0,t_0+T-\delta)$ and instances fully appearing in $[t_0+T-\delta,t_0+T)$.

Similar to the previous section, we need to 1) compute the number of possible isomorphic subgraphs to the motif graph $G_\bm$; and 2) calculate the probability that timestamps of the edges in each isomorphic subgraph occur in the order $\prec_M$ specified by the motif within a time window of length $\delta$. 

In the sequel, we explain the detailed steps in deriving the following Theorem.
\begin{theorem}\label{thm:2}
	The expected number of $\delta$-instances of a temporal motif $\bm$ in a Temporal Activity State Block Model (\alg) during a time interval $[t_0, t_0+T)$ for $T > \delta$ is
\begin{align*}%\label{eq:cnt_expbigint}
\E [N_\bm|T> \delta] = \E[N_\bm|T=\delta] +
 \E [N_\bm | T\!>\!\delta, t_1\!<\!t_0+T\!-\!\delta],
\end{align*}
where $t_0$ and $t_1$ are the timestamps of the first and second edges of the motif instance.
\end{theorem}
\xhdr{Expected Number of Isomorphic Subgraphs}
To calculate the number of motif instances with the first edge occurring in $[t_0,t_0\!+\!T\!-\delta)$, 
assume that $e'_1=(u',\!v') \!\in\! E_\bm$ is the edge of the motif that comes first in the ordering $\prec_M$. 
We first compute the number of isomorphic 2-nodes subgraphs in the  network to $\{u',v'\}$. Then, for each isomorphic subgraph, we calculate the number of subgraphs in the remaining network isomorphic to $V_\bm \setminus \{u',v'\!\}$. 
%the two subgraphs $G_\bm^1=(V_\bm^1, E_\bm^1)$, where $V_\bm^1=\{u',v'\}, E_\bm^1=\{e'_1\}$, and $G_\bm^2=(V_\bm^2, E_\bm^2)$, where $V_\bm^2= V_\bm \setminus \{u',v'\}, E_\bm^2=E_\bm \setminus e'_1$.

%%%There are $B_{n,2,C}$ unique bijections $f_1:\{u,v\} \rightarrow \{u',v'\}$ from 2-node subgraphs in the network to $E_\bm^1$. For each bijection $f_1:\{u,v\} \rightarrow \{u',v'\}$, there are a set of %${n-2 \choose k-2}$ ways to form $(k-2)$-node subgraphs from $V \setminus \{u,v\}$. For each permutation of the $k-2$ nodes we get a unique bijection. $|B_{n-2,k-2,C}|$ bijections $f_2: V_r \rightarrow V_\bm \setminus \{u',v'\}$, where $V_r$ is a set of $k-2$-nodes from $V \setminus \{u,v\}$. %from the nodes of the network that are not selection to $$.
%%
%%To count the expected number of isomorphic subgraphs resulted from $f=(f_1, f_2)$, we calculate the expected number of times $(u,v)$ may occur in $[t_0,t_0+T-\delta)$. For each occurrence, we calculate the expected number of times the $z-1$ remaining edges may appear in $[t_1, t_1+\delta)$, where $t_1$ is the time of appearance of the first edge.
%
\begin{algorithm}[t]
	\caption{Calculate Expected Motif Frequency}
	\label{alg:atsbm}
	\begin{algorithmic}[1]
		\Require Set of nodes $V$, set of $C$ activity states, edge arrival rates between activity states $\theta_{ij}$, time interval $[t_0,t_0+T)$. %, \alg~ instance with $C$ groups
		\Ensure Expected number of $\delta$-instances of motif $\bm$ within time interval $[t_0,t_0+T)$.
		
		\For{$(i, j) \in ([C^{out}]\times[C^{in}])$}
		\State $E[N_{e_{ij}}]$ = \Call{Integrate}{$\theta_{ij}$,~$t_0$,~$t_0+T$}
		\EndFor
		
		\State $\E[N_\bm] = 0$
		\State $\mathcal{A}_{C,k} = $ set of $O(C^k)$ activity state assignments to $k$ nodes. %from $C$ groups.
		
		\For {$A \in \mathcal{A}_{C,k}$}
		\State $|\mathcal{B}_{k,V,C}|=\prod_{c\in [C]} P(n^c, n^c_s)$ \Comment{Eq. \ref{eq:group_count}}
		\State $V_s=\{v_1, \cdots, v_k |a_{v_i}=A[i] \}$ %a set of $k$ nodes consistent with activity states $A$.
%		\State $f:V_s \to V_{\bm}$ such that $f(e_1) \prec_M f(e_2) \prec_M \cdots \prec_M f(e_z)$
		\State $f:$ a bijection from $V_s$ to $V_\bm$ 
		\State $\E[N_{S}] = 1$
		\For{$(u,v) \in V_s$ such that $(f(u), f(v)) \in E_\bm$} \Comment{Eq. \ref{eq:dyn_num}}
		\State $\E[N_{S}] = \E[N_{S}] \times  E[N_{e_{a_ua_v}}]$ 
		\EndFor
		
		\State $P_{\text{order}} =Pr(t_0\leq t_1<t_2<...<t_z< t_0+T)$ \Comment{ Lemma \ref{lemma2}}
		\State $\E[N_{\bm}] = \E[N_{\bm}] + |\mathcal{B}_{k,V,C}| \times \E[N_{S}] \times P_{\text{order}}$ %\Comment{Eq. \ref{eq:group_count}} 
		
%		\State $P_{\text{order}} =1$
		
%		\For{$i = z, z-1, \cdots, 1$} %\Comment{Eq. \ref{eq:group_count}} 
%			\State $\Theta_{e_{a_ia_j}}= \theta_{ij} / $ \Call{Integrate}{$\theta_{ij}$,~$t_{i}$,~$t_0+T$}		
%			\State $P_{\text{order}} = P_{\text{order}} \times$ \Call{Integrate}{$\Theta_{ij}$,~$t_{i}$,~$t_0+T$}			
%		\EndFor
		
		%\State $\E[N_\bm] = \E[N_\bm] + \E[N_{S}] \times P_{\text{order}}$
		\EndFor
		
		\hspace{-.5cm}\Return $\E[N_\bm] $

	\end{algorithmic}
\end{algorithm}

\begin{lemma}\label{lemma3}
	The expected number of $k$-node subgraphs isomorphic to the motif subgraph $G_\bm$  in a Temporal Activity State Block Model (TASBM) with the first edge occurring in $[t_0, t_0+T-\delta)$ for $T > \delta$ is
\begin{align*}%\label{eq:dyn_exp2}
%\E [N_\bm|&T>\delta, t_1<T-\delta] = \\
\E [N_{S^{k,z}_{V,C}} &\mid t_1 \in [t_0,t_0\!+\!T\!-\!\delta)] = \\
%&\sum_{S_{k,z}} \int_0^{T-\delta} \theta_{e_1}(t) dt \cdot   \int_0^{T-\delta} \Theta^{[0,T-\delta]}_{e_1}(t_1)\cdot 
%\left[ \prod_{i =2}^z \int_{t_1}^{t_1+\delta} \theta_{e_i}(t)dt \right]dt_1  \nonumber\\
%& \sum_{\substack{f \in B_k, \\f: V_s \rightarrow V_\bm}}  
%& |\mathcal{B}_{n,2,C}| \int_{t_0}^{t_0+T-\delta} \theta^{[t_0,t_0+T-\delta]}_{a_ua_v}(t_1) \cdot \nonumber\\
%|\mathcal{B}_{n-2,k-2,C}|  \prod_{i =2}^z \int_{t_1}^{t_1+\delta} \theta_{a_sa_r}(t)dt  dt_1.  \nonumber
& \E[N_{S^{2,1}_{V,C}}|{t\in[t_0, t_0+T-\delta)}] \cdot \E[N_{S^{k-2,z-1}_{V\setminus V_{e_1},C}}|{t\in[t_1, t_1+\delta)}], \nonumber
\end{align*}
where $t_1$ is the time of appearance of the first edge, and $V_{e_1}$ is the set of two nodes in subgraph $S^{2,1}_{V,C}$.
\end{lemma}

\xhdr{Probability of Temporal Ordering for \pmb{$T>\delta$}}
We now use marginal edge probabilities to calculate expected frequency on intervals with $T>\delta$. For subgraph $G_S$ isomorphic to $G_M$, we compute the probability that the first edge occurs in $[t_0,t_0+T-\delta]$ and the remaining $z - 1$ edges occur sequentially within a time window of length $\delta$ starting at $t_1$.
%it is a motif instance, to be used in summing over all possible vertex bijections similar to in Eq.~\ref{eq:exp_dyn}:
\iffalse
\begin{align*}%\label{eq:dyn_t}
\E [N_\bm&|T>\delta, t_1\!<\!t_0\!+\!T\!-\!\delta] = \E[N_{S_{k,z}} \!\mid \! t_1 \in [t_0,t_0+T-\delta)] \cdot \\
& Pr (t_2<t_3<...<t_z< t_1+\delta | t_1 \!<\! t_2, t_1 \in [t_0,t_0\!+\!T\!-\!\delta))\nonumber
\end{align*}
with:\fi

\begin{lemma}\label{lemma4}
	The probability that temporal edges of a subgraph $G(V_s, E_s)$ occur in the order $\!\prec_\bm\!$ specified by motif $\bm$ in an interval of length $T\!\!>\!\!\delta\!$ with the first edge appearing in $[t_0, t_0\!+\!T\!\!-\!\!\delta)$  is
\begin{align*}%\label{eq:dyn_prob}
Pr &(t_2<t_3<...<t_z< t_1+\delta | t_1 \!<\! t_2, t_1 \in [t_0,t_0\!+\!T\!-\!\delta)) = \\ 
&\int_{t_0}^{t_0+T-\delta} \!\!\! \Theta^{[t_0,t_0+T-\delta]}_{e_1}(t_1) \int _{t_1}^{t_1+\delta} \!\!\! \Theta^{[t_1,t_1+\delta]}_{e_2}(t)  \int_{t_2}^{t_1+\delta} \!\!\! \Theta^{[t_1,t_1+\delta]}_{e_3}(t_3) \nonumber\\ 
&\hspace{.5cm} \cdots \int_{t_{z-1}}^{t_1+\delta} \Theta^{[t_1,t_1+\delta]}_{e_z}(t_z) dt_{z} dt_{z-1} dt_{z-2}...dt dt_1. \nonumber
\end{align*}
\end{lemma}

The expected frequency of $\delta$-instances of temporal motif $\bm$ in a time window of length $T > \delta$, conditional on at least one edge occurring in $[t_0, t_0\!+\!T\!-\!\delta)$, can be calculated using Lemmas \ref{lemma3} and \ref{lemma4}.
\begin{align}\label{eq:dyn_exp3}
\E [N_\bm&|T>\delta, t_1\!<\!t_0\!+\!T\!-\!\delta] = \E[N_{S^{k,z}_{V,C}} \!\mid \! t_1 \in [t_0,t_0+T-\delta)] \cdot \\
& Pr (t_2<t_3<...<t_z< t_1+\delta | t_1 \!<\! t_2, t_1 \in [t_0,t_0\!+\!T\!-\!\delta))\nonumber
\end{align}

Finally, The expected number of instances fully appearing in $[t_0+T-\delta,t_0+T)$ can be calculated from Theorem \ref{thm1-1}. Hence, we get Theorem \ref{thm:2} for the expected number of $\delta$-instances of temporal motif $\bm$ in a time interval $T > \delta$. %using the following Theorem. %\E [N_\bm|T> \delta] $ for the block model:

\subsection{Variance of motif counts}\label{sec:var}
We next discuss how we can use our framework for deriving the variance of the number of motif instances $\mathbb{V}[N_\bm]$. %can be reduced to computed expected frequency of additional motifs.
I.e., $$\mathbb{V}[N_\bm] = \E[N_\bm^2]-\E[N_\bm]^2.$$
While we can simply calculate $\E[N_\bm]^2$ using Algorithm \ref{alg:atsbm},
computing $\E[N_\bm^2]$ involves calculating the expected number of \textit{pairs} of $\delta$-instances of motif $\bm$. A pair of instances can overlap in up to $k$ vertices and up to $z$ temporal edges. %but may also be completely disjoint. 
%Examples of two overlapping motif instances sharing vertices and/or edges are show in Appendix Figure~\ref{fig:var}.

In order to calculate $\E[N_\bm^2]$, we need to consider both independent and dependent pairs of $\delta$-instances of $\bm$.
Motifs instances which do not share an edge (see Appendix Figures~\ref{fig:shared1} and \ref{fig:shared2} for examples) are conditionally independent. On the other hand, pairs of instances which share at least one edge are not independent. %
%Hence, we need to take into account subgraphs of up to $2k$ nodes and $2z$ edges.
 %$$ \E[N_\bm^2] = E[N_{(G_{s_1}, G_{s_2})| E_{\bm_1} \cap E_{\bm_2} = \emptyset}] + E[N_{(\bm_1, \bm_2)| E_{\bm_1} \cap E_{\bm_2} \neq \emptyset}]$$
%Thus we compute $\E[N_M^2]$ as the expected number of disjoint pairs of instances of $M$ plus the expected number of intersecting instances of $M$. 
For dependent instances, we must consider all possible total orderings of the edges of the two pairs of $\delta$-instances of $\bm$. 
For example there is a pair of $\delta$-instances such that $t_1<t_2<t_3 = t_{3'}$ from the first $\delta$-instance and $t_{1'}<t_{2'}<t_{3'}=t_{3}$ from the second $\delta$-instance, but the ordering of the edges $\{t_1,t_2,t_{1'},t_{2'}\}$ is not fully specified by the two instances and thus we need to calculate all the possibilities for the remaining 4 edges, including $t_1<t_2<t_{1'}<t_{2'}$, $t_{1}<t_{1'}<t_2<t_{2'}$, $t_{1'}<t_{2'}<t_1<t_{2}$, etc (see Appendix Figure~\ref{fig:shared3}).
 
% Let $\mathcal{M}$ be the set of pairs of $\delta$-instances of $\bm$ that can be constructed by combining two $\delta$-instances of $\bm$, and their time intervals.
 Let $S_1, S_2$ be a pair of of $\delta$-instances of $\bm$. The time interval %$\delta_{(S_1,S_2)}$ of 
 that the edges $E_{S_1}\cup E_{S_2}$ of both instances may occur is within an interval $[t_0+\delta, t_0+2\delta)$, depending on which temporal edges, if any, are shared. We denote by $\E_\delta$ and $\E_{\delta'}$ expected $\delta$- and $\delta'$-instances of motif $\bm$, respectively. Then by linearity of expectation, we get
 %
 %the expected number of dependent pairs of $M$ is $\sum_{M' \in \mathcal{M}} \E[M']$. Thus variance can be simply computed by applying our analysis to larger motifs. %, but is computationally more demanding as it requires counting larger motifs which must considered.
 %
\begin{align*}
\E[N_\bm^2] =\hspace{-.4cm}\sum_{\substack{(S_1,S_2):\\E_{S_1}\cap E_{S_2} = \emptyset}}\hspace{-.4cm}\E_\delta[N_{S_1}|t\in[t_0,t_0+T])\E_\delta[N_{S_2}|t\in[t_0,t_0\!+\!T)]+\\\sum_{\substack{(S_1,S_2):\\E_{S_1}\cap E{S_2} \neq \emptyset}}\sum_{\delta'\in[t_0+\delta,t_0+2\delta)}\E_{\delta'}[N_{S_1\cup S_2}|t \in[t_0,t_0\!+\!T) ].
%&= \hspace{-.8cm}\sum_{t({E_{S_1}}) \cup t({E_{S_2}}) \in [\delta, 2\delta)} \hspace{-.8cm} E[N_{({S_1}, {S_2})| E_{S_1} \cap E_{S_2} = \emptyset}] + \hspace{-.8cm} \sum_{\substack{ E_{S_1} \cap E_{S_2} \neq \emptyset}} \!\!\!\!\E[N_{(S_1,S_2)}].
\end{align*}
 
  %Consider the three possible cases for a super-motif $M'$ made by combining two instances of a $3-$vertex $3$-edge motif $M$: $M'$ may have $6$, $5$, or $4$ edges, depending on if $0$, $1$, or $2$ edges are shared between the two instances of $M$. Since subgraphs isomorphic to overlapping instances of $M$ may also have multiple valid temporal orders, we must compute the expected number of all $5$-vertex $6$-edge, $4$-vertex $\{5,4\}$-edge, and $3$-vertex $\{6,5,4\}$-edge motifs to add up the expected number of all possible instances of $M'$.

\iffalse
\begin{figure}[t]
	\centering
	\subfloat[\label{fig:shared1}]{\includegraphics[width=.25\textwidth]{FIG/motif_overlaps_a.pdf}} 
		\subfloat[\label{fig:shared2}]{\includegraphics[width=.21\textwidth]{FIG/motif_overlaps_b.pdf}} 
			\\\vspace{-3mm}
		\subfloat[\label{fig:shared3}]{\includegraphics[width=.18\textwidth]{FIG/motif_overlaps_c.pdf}} 
		\subfloat[\label{}]{\includegraphics[width=.14\textwidth]{FIG/motif_overlaps_d.pdf}} 
	\vspace{-2mm}
	\caption{Examples of joint instances of motif $M$: $S_1$ on vertices $(a,b,c)$ with edges labeled $1,2$, and $3$ at times $t_1<t_2<t_3$ and $S_2$ on vertices $(d,e,f)$ with edges labeled $1',2'$, and $3'$ at times $t_{1'}<t_{2'}<t_{3'}$. }
	\label{fig:var}
	\vspace{-.4cm}
\end{figure}\label{fig:motifoverlap}
\fi

\subsection{Computational Complexity}
Here, we derive the computational complexity of our method to compute the expected number of $\delta$-instances of a $k$-node $z$-edge temporal motif in a \alg~ with $C$ activity states.
The first for loop (line 2) in Alg. \ref{alg:atsbm} calculates $C^2$ integrals to get the expected number of edges between every pair of activity states. 
For each of the $|\mathcal{A}_{C,k}| = O(C^k)$ activity state assignments (line 5), we iterate over $z$ edges (line 10) to calculate the expected number of isomorphic subgraphs. Then we calculate the probability of the correct temporal ordering for the edges of the isomorphic subgraphs (line 12). 
If edge arrival rates between activity states do not change within the time interval, the probability of correct ordering is a constant $(1/z!)$. 
In the general case, this involves calculating a set of $z$ integrals as specified in Lemma \ref{lemma2}.
Assuming the cost of computing each integral is $O(1)$, the total computational complexity of Algorithm \ref{alg:atsbm} is $O(C^k)$. 
For the case where $t>\delta$, the additional computation of $\mathbb{E}[N_M|T>\delta, t_1\!<\!t_0\!+T\!-\delta]$ for $T>\delta$ (Eq.~\ref{eq:dyn_exp3}) follows the same structure as in the case where $T \leq \delta$ with different integral bounds.
Therefore, the computational complexity for calculating the expected motif frequencies is $O(C^k)$, where $C$ is the number of blocks (usually $<10$) and $k$ is motif size (usually $<10$). %and thus this is the complexity of the method on any interval $[0,T]$. 

Notice the number of blocks $C$ and the motif size $k$ are constant relative to the size of the network, we get a computational complexity of motif counting to be $O(1)$. Due to its low computational complexity, our analytical method can easily scale to large real-world temporal networks with millions of temporal edges. 

%Our method and analysis are easily generalized to any temporal network model by considering $C=n$ activity states, one for each node. However, without benefiting from \alg, we have $C = |V|$, and the computational complexity of calculating the expected motif frequencies is $O(n^k)$. %, where $k$ is the number of bijections between the nodes/edge of a motif.

%We next compare these to edge re-wiring and timestamp shuffling approaches. In either such randomized ensemble technique, suppose $r$ trials are generated: then $r$ temporal motif enumerations must be done 
%Furthermore, note that in edge re-wiring and timestamp shuffling approaches, for an ensemble of size $r$, we need to generate $r$ (in the order of thousands) randomized networks and enumerate the motif instances on each of them to calculate the average and standard deviation for the number of instances of each motif. 
%While even creating thousands of randomized networks is computationally expensive, each enumeration of motifs on an instance takes at least $O(|E|)$ steps (see~\cite{paranjape2017motifs} for algorithms) and thus the $O(r|E|)$ cost of motif enumeration in an ensemble is much greater than cost for analytical method using  \alg.

\iffalse
 In practice, we also use constant edge arrival rates with the stochastic block model over sub intervals of the data. When rate changes, the node changes groups, and we recalculate groups at a fine enough interval to catch this so we don't have to also have dynamic rate within an interval.
Also: precompute/reusing when groups change but inner product/sum doesn't
\fi

\section{Experiments}
\label{sec:experiments}
% !TEX root = motifs_newarxiv.tex

%\jure{I think it is imporant to keep mentioning expected count and variance everywhere and then only in section 6 we only mention expected (and no variance). Only here we then acknowledge we only compute expectation due to computational complexity.}

\begin{figure}
	\includegraphics[width=0.6\textwidth,height=0.6\textwidth]{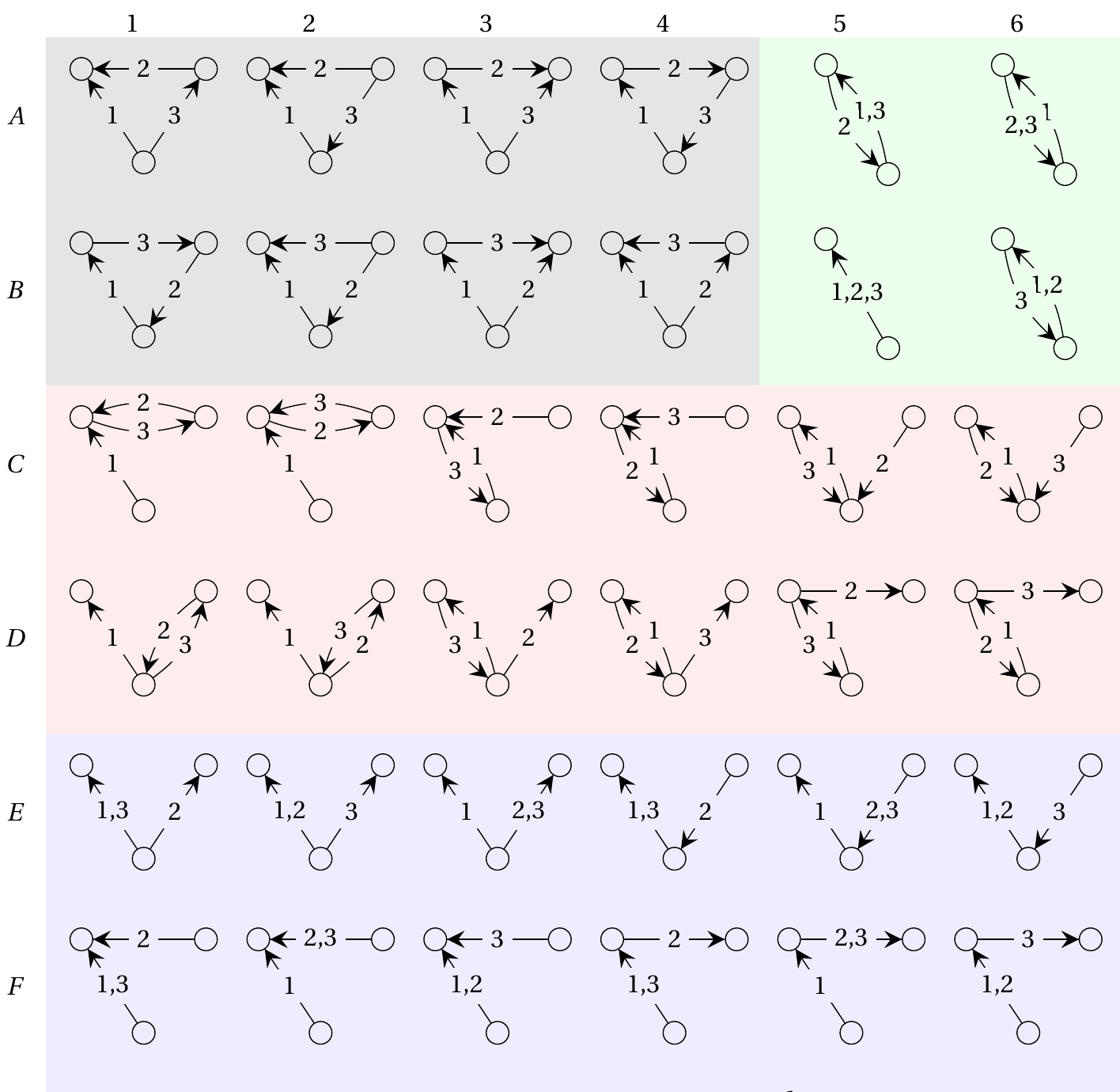}\vspace{-.12cm}
	\centering\caption{All 2- and 3- node motifs with 3 edges, shaded by main structural feature:  triangles (A1-4, B1-4; grey), Two-Node (A5,6 B5,6; green), Reciprocated edge (C1-6,D1-6; red), and Double edge (E1-6,F1-6; blue).}\label{fig:motifgrid}
	\vspace{-.5cm}
\end{figure}

In this section, we present the results of applying our analytical framework to an inferred \alg~ to calculate the expected motif frequencies in synthetic and real-world temporal graphs. We compare the expected number of motif instances to the number of observed motif instances counted by the method of \cite{paranjape2017motifs}. 
To demonstrate our \alg~ model and analytical method, we focus on expected counts for motifs with 2 or 3 nodes and 3 edges (Figure~\ref{fig:motifgrid}). 

%While our framework can be used for larger motifs, existing algorithm for counting temporal motif frequencies do not scale well to motifs with larger than 3 edges. Therefore, we only report the results for 3-edge motifs with 2 and 3 nodes, as shown in Figure~\ref{fig:motifgrid}, and do not include experiments computing variance.

%including those representing joint motif instance (e.g., Figure~\ref{fig:var}), we do not compute these in practice due to the computational cost. Thus we do not include experiments computing variance, which require such counts described in Section~\ref{sec:var}.

%Although \cite{paranjape2017motifs} is the most efficient existing algorithm for counting temporal motif frequencies, it does not scale well to motifs with larger than 3 edges. Therefore, we only report the results for 3-edge motifs with 2 and 3 nodes, as shown in Figure~\ref{fig:motifgrid}. We note that our analytical framework can be used to calculate the expected frequencies for larger motifs.

We first discuss how we fit the \alg~ model to observed temporal network data. We then present experiments which show the effect of network properties and model parameters, i.e., average degree of the temporal network, length of the time window $T$, and motif window $\delta$ on how well our model estimates motif frequencies in synthetic data. Next, we show that our analytical framework can accurately track motif frequencies in real-world networks, including a financial transaction network, an email network, and a phone call network. Finally, we show how our model can be used for anomaly detection, by identifying places where the observed counts deviate significantly from the model.

\subsection{Fitting the \alg Model}\label{sec:fit}
Our framework first computes the average out-edge and in-edge arrival rates for all nodes on every window of $T$ time units $[iT\!,(i+\!1)T)$, $i \geq 0$. Out-edge rates are partitioned into $C^{out}$ groups and in-edge rates are partitioned into $C^{in}$ groups, for a total of $C\!= \!C^{in}\!\cdot C^{out}$ groups corresponding to the out- $\!$and in- $\!$edge rate combinations. In a single pass over the edges of time interval $[0,T]$, we can determine the out- and in-degree of each node. To assign our group partitions in a single pass over the nodes at the end of the time interval, we take the following approach. We start by partitioning the full range of possible edge arrival rates, and assign each range to a group. Initially, there are many such groups, i.e. a large value of $C^{in}$ and $C^{out}$.  Then nodes are assigned to the group which corresponds to the range containing their observed rate, for out- and in-rates independently.  Empty groups can be dropped from subsequent computations, so starting with a large initial set does not incur significant computational cost.

We then use our group assignments to approximate $\pmb{\theta}$ without making another pass over the edges. During assignment of nodes to groups, we compute the observed average out- and in- degrees of each group in $[C^{out}]$ and $[C^{in}]$, respectively. These total out and in rates correspond to the column and row sums of $\pmb{\theta}$. Rather than iterating over all edges once group assignments have been made to determine the exact breakdown of each sum, we make the assumption that the out-edges of a group $i \in [C^{out}]$ will be distributed among all nodes according to their in-edge rate. Thus we need only the set of average in-rates for each group in $[C^{in}]$ to determine how the total rate for $i \in [C^{out}]$ is divided up over a row of $\pmb{\theta}$. In practice, this method results in an accurate approximation of $\pmb{\theta}$, so we use it for all the following experiments.

%The nodes are then partitioned into $C$ groups and expected motif frequencies are computed using the average arrival rates for the nodes in each group. 

%Our framework takes the length of the time window $T$ as input and computes the average edge arrival rates of all temporal edges on every interval $[iT,(i+1)T), i \geq 0$. The other inputs to the implementation are the motif window, $\delta$, and the maximum allowed number of vertex groups, $C$. 
%Each group $c \in [C]$ has an initial value of arrival rates is given an initialization value and then vertices are assigned to the group with a value closest to their observed rate. Expected motif frequencies are then computed using the average edge rates actually observed for the vertices assigned to each group, rather than the initialization value and empty groups are dropped to reduce the computational complexity. While we could use the observed rates from the first interval to determine group initialization values, this requires non-trivial computation, and does not have a significant impact on the resulting partition. %\textbf{[why don't we calculate the rates on the firs interval and divide the nodes to C groups based on that?]}

%\jure{I find the notation $C=C$ strange. Why don't you invent a symbol for the number of groups (say $k$)and then use that to say how many groups are there.}

%\textbf{[explain the grouping you used to report the synthetic results: triangles, 2-vertex, ...]}

\subsection{Accuracy on Synthetic Networks}\label{sec:syntheticresults}
%We first test our motif frequency null model on synthetic temporal networks. 
Our synthetic networks are generated according to the TASBM with $C\!=\!C^{out}\!\cdot \!C^{in}$ groups with $C^{out}$ out-edge and $C^{in}$ in-edge states, respectively. Each node is assigned to a  group in $[C]$ chosen uniformly at random.
Each pair of groups $(C_1,C_2)$ is also assigned an arrival rate, specifying the arrival rate of edges $(u,v)$ such that $u \in C_1$ and $v \in C_2$. For each pair of nodes $(u,v)$, temporal edges are then sampled according to a Poisson process with the rate corresponding to the out-edge group of $u$ and in-edge group of $v$.

\begin{table}[!t]%\vspace{-.3cm}
	\small
	\centering
	\begin{center}\vspace{-.2cm}
		\begin{tabular}{||c c c c c||} 
			\hline
			C & Triangles & Two Vertex & Reciprocated & Double Edge \\ [0.5ex] 
			\hline \hline
			1 &  0.229 &  0.381 &  0.147 & 0.381\\ \hline
			4 & 1.99e-05 & 4.35e-05 &2.84e-05 &1.69e-05\\ \hline
			9 &  1.89e-05 &  4.26e-05 & 2.78e-05 & 1.60e-05 \\ \hline
			16 &  1.04e-05 & 3.59e-05 &2.25e-05& 7.90e-06 \\ \hline
			25 &  1.04e-05&  3.59e-05 & 2.25e-05 & 7.91e-06\\
			\hline\hline
		\end{tabular}
	\end{center}
	\caption{MSRE for varying number of groups $C$ used by \alg~ using $T \!=\! 10K$, $\delta \!=\! 5K$. The values are calculated over $30$ networks generated with $300$ nodes and $C^{out}\!=\!C^{in}\!=\!5$.}\label{table:groupcount}
	\vspace{-.6cm}
\end{table}

\xhdr{Accuracy of Model Inference}
We first show that we can accurately infer the \alg~ parameters used to generate the synthetic network and thus accurately use our analytical approach to determine expected motifs. We measure accuracy of our  model over a set of $r$ synthetic networks, using Mean Squared Relative Error, $$MSRE = \frac{1}{r} \sum_{i=1}^r \left(\frac{N^i_{\bm}-N_\bm}{N^i_{\bm}}\right)^2,$$
where $N^i_\bm$ is the actual number of motif instances counted using the method of \cite{paranjape2017motifs} in the $i$-th generated network, and $N_\bm$ is the expected motif frequency calculated by our framework.

%comparing the  expected motif frequencies calculated using our framework to the actual number of motif instances counted using the method of \cite{paranjape2017motifs}, using %in a number of synthetic graphs using 
%Mean Squared Relative Error: $$MSRE = \frac{1}{n} \sum_{i=1}^n \left(\frac{N^i_{\bm}-N_\bm}{N^i_{\bm}}\right)^2,$$ where $N^a_\bm$ is actual motif frequency in the $i$-th generated graph, and $N_\bm$ is the expected motif frequency calculated by our framework. %%Since the number of instances of different motifs may be significantly different, we use SMSRE to make the results comparable.

%\vspace{.2cm}
%\textbf{Accuracy for varying number of groups in block model.}\label{sec:synthgroupcount}
 Table \ref{table:groupcount} shows MSRE for 30 networks with 300 nodes generated using \alg~ with $C^{out}\!=\!C^{in}$. 
For generating out-edges, we partition the nodes to groups of size 10, 30, 60, 80, and 120, with out-rates of 1e-7, 1e-6, 1e-5, 1e-4, and 1e-3.
For generating in-edges, we have all the nodes in one group, hence having in-rate of 11111e-3.
%We use 20 empty groups and one group of each of 10, 30, 60, 80, and 120 nodes with total edge rates of 1e-7, 1e-6, 1e-5, 1e-4, and 1e-3 from each of the five non-empty groups, respectively, to all other non-empty groups (including itself). 
Generated networks have an average of 384,580 edges.
We varied the number of groups in our framework for calculating the expected motif frequencies from $C^{out}\!=\!C^{in}\!=\!1$ to $C^{out}\!=\!C^{in}\!=\!5$ and used $T\!=\!10K$ and $\delta\!=\!5K$ time units. It can be seen that the error quickly vanishes %reduces to zero %expected motif frequencies get closer to the observed motif counts 
when the model is allowed to use a higher number of groups. However, the improvements from increasing the number of groups quickly diminish. %While the number of groups used to generate the synthetic graphs for the experiment in Table~\ref{table:groupcount} was $C_g = 5$, 
It can be observed that using only $C^{out}\!=\!C^{in}\!=\!2$ groups to calculate the expected frequencies, we get almost the same accuracy as using $C^{out}\!=\!C^{in}\!=\!5$ groups. This indicates that while real data is likely to have a large variety of node activity levels, a relatively small value of $C$ can be used to calculate accurate expected motif counts.
%\textbf{[in the caption of the table you have 30 groups?]}

\begin{figure*}[t]%\vspace{-.4cm}
	%\centering  
	\subfloat[{\label{fig:msre_r}}]{\includegraphics[width=.48\textwidth,height=.28\textwidth]{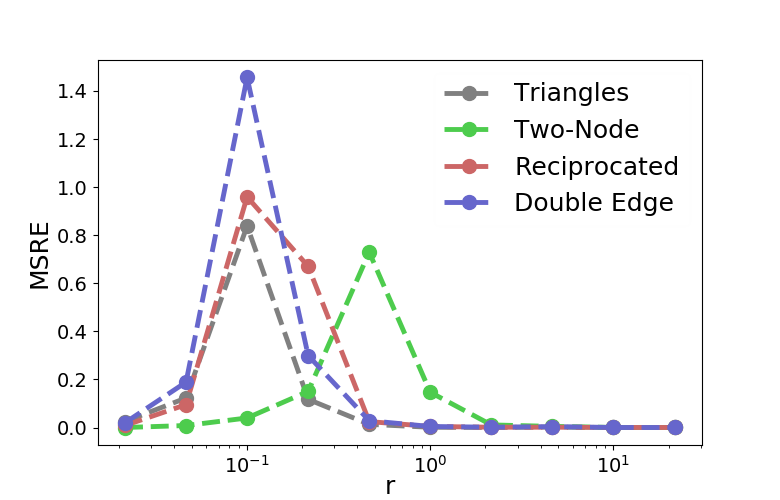}}\hspace{0.2cm}
	\subfloat[{\label{fig:msre_d}}]{\includegraphics[width=.48\textwidth,height=.28\textwidth]{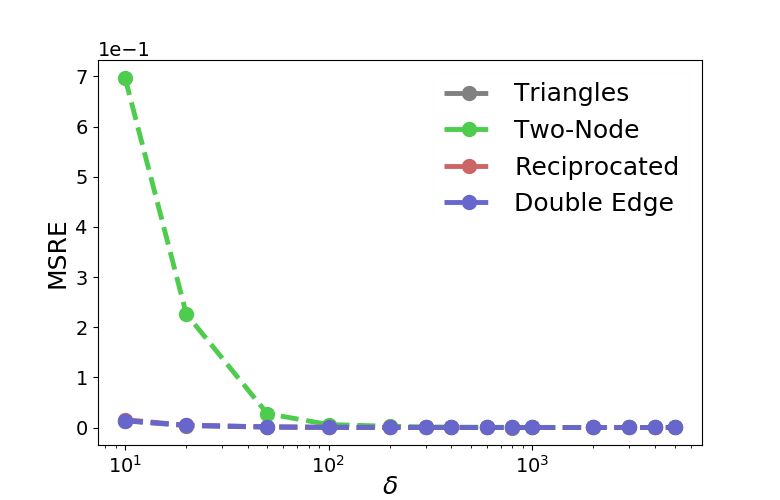}}\\
	\subfloat[{\label{fig:msre_t}}]{\includegraphics[width=.48\textwidth,height=.28\textwidth]{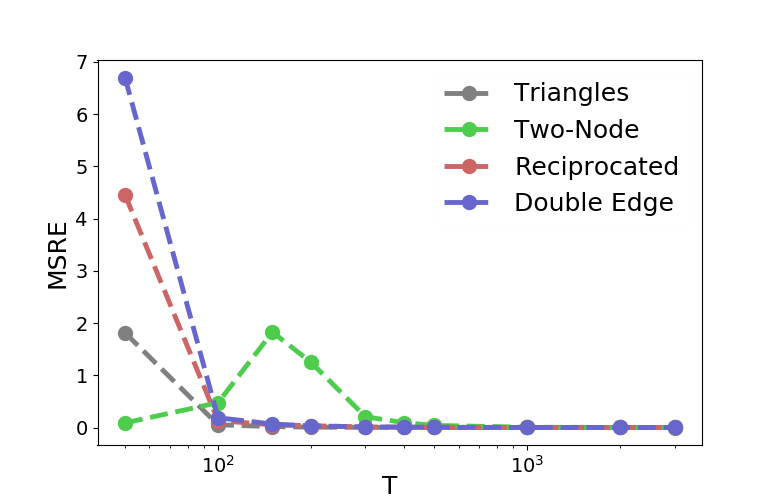}}\hspace{0.2cm}
	%\subfloat[{\label{fig:msre_i}}]{\includegraphics[width=.22\textwidth,height=.18\textwidth]{FIG/allmotifsoverint_msre.png}}
	%	%\centering
	\vspace{-.4cm}
	\caption{MSRE for the expected motif counts over 30 synthetic networks with $100$ nodes for varying (a) number of edges, (b) motif window $\delta$, and (c) time window $T$. See Figure~\ref{fig:motifgrid} for motif names. Notice that the TASBM accurately predicts motif counts. % and d) SRE for varying number of intervals, for a network with $100$ nodes generated over 64 intervals of length $10K$, with distinct edge arrival rates per interval. %scale $r$.
	%[Use a larger font in the figures, you can also drop the titles. We used nodes in the paper, so probably you can change the legend to twoNodes.]
	}\label{fig:synthvars} 
	\vspace{-.4cm}
\end{figure*}

%\vspace{.2cm}
\xhdr{Robustness of Model to Hyper-Parameter Choices}
Next we investigate robustness of our method to choices of hyper-parameters.
Figure \ref{fig:synthvars} compares MSRE for 30 networks with 100 nodes generated using \alg~ with $C^{out}\!=\!C^{in}\!=\!3$. %, and a set of distinct but fix edge arrival rates between groups. \textbf{[what are the values?]}
Here, for generating out-links we divide the nodes into groups of sizes 10, 30, and 60 %(along with 6 empty groups) 
and use the initial out-rates of 5e-6, 1e-4, and 1e-3, respectively. %, from each non-empty group to all other non-empty groups (including itself). 
The rates were chosen to be sufficiently distinct and to generate sufficiently large edge volumes without motif counts exceeding the maximum capacity of the motif counter from~\cite{paranjape2017motifs}. 

%We calculated the expected motif frequencies using $C = 3$, and varying the time window $T$, the motif window $\delta$ and edge arrival rates between groups.
%For the remaining experiments, we use a set of 30 synthetic graphs generated with $100$ nodes divided into $C_r=3$ distinct groups, and $C = 3$. 
%, as expected, in this setting the model
%It can be seen that the model becomes more accurate as the time window $T$ increases, the motif window $\delta$ increases, or the total edge rate increases. The plots show the mean squared relative error (MSRE) for 30 different synthetic graphs generated with the same randomly assigned stochastic block model varying each parameter.

%Figure~\ref{fig:msre_r} shows that the model becomes more accurate as the time window $T$ increases, the motif window $\delta$ increases, or the total edge rate increases.

Figure~\ref{fig:msre_r} shows the accuracy of our model for networks with increasing average degree, generated by scaling the entire set of arrival rates exponentially, both within and between groups. %, by a factor of $r$. 
The peaks in error as the average degree increases occur at approximately the point when motif counts become non-zero, but have high variance due to low edge rates.
For  two-node motifs, this peak 
%This peak in MSRE before convergence to zero 
happens at a larger edge volume since the frequency of such motifs are lower than that of three-node motifs.

Figure~\ref{fig:msre_d} shows how MSRE converges to zero as the motif window $\delta$ increases. %with a fixed window size of $T=10000$ and rate factor of $r = 1$, using the same block model and rates as for Figure~\ref{fig:msre_r}. \textbf{[what is a time step here? do you generate 1 edge per timestep?]}. 
The convergence is due to the same reason as above; as the motif window increases there is a greater volume of randomly generated edges over which each computation is made, and thus the average motif counts are closer to our expected counts. 
%\textbf{[what is $\delta$ here? As $T>\delta$ for larger $\delta$, T is larger and the arrival rates are more accurate.]}
Again, as there is a smaller number of two-node motifs in a temporal network generated by stochastic block model, the MSRE converges more slowly for two-node motifs.

Figure~\ref{fig:msre_t} shows a similar behavior for increasing the length of the time window $T$, for a fixed motif window of $\delta = 5K$. 

\subsection{Accuracy on Real-world Networks}\label{sec:realworldresults}
In our real-world experiments, we apply our analytical model to a financial transaction network and an email network. We show that our model matches the trends in motif counts in the real data. Since we are interested in modeling temporal dynamics of the networks, we preprocess the financial transaction and email networks by removing low-degree ($<10\%$ the largest degree) nodes and focusing on the largest connected community. Due to space constraints we present results for a subset of motifs. Full results can be found in the Appendix.
%We also show that the scale of the model's motif counts are more accurate when applied to a subset of the network: the largest community formed by nodes with at least 10\% of the maximum degree. Here we show selected  motifs representative of the full set. Results for the 36 motifs shown in Figure~\ref{fig:motifgrid} can be found in the Appendix.
%\vspace{.2cm}

\begin{figure} 
	%\subfloat[{\label{fig:bankplots}}]{\includegraphics[width=0.42\textwidth,height=.14\textwidth]{FIG/KDDversions/bank_both_00_41.png}}\\
	%\vspace{-.3cm}
	%\subfloat[{\label{fig:bankplotspreproc}}]{
	\includegraphics[width=0.84\textwidth,height=.28\textwidth]{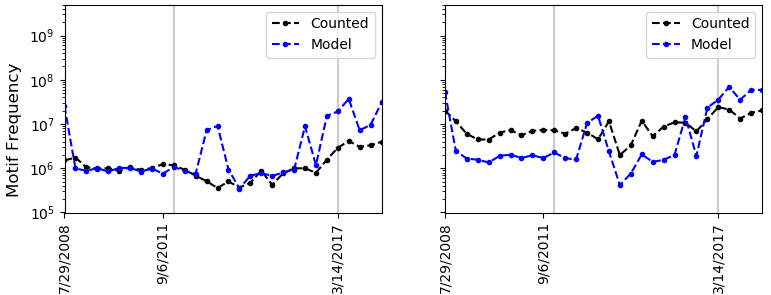}
	%}
	\vspace{-.3cm}
	\caption{Financial transaction network, $\delta\!=\!T\!=\!$ 90 days. Motifs F1 and A6. %(a) Financial transaction network, (b) low-degree nodes ($<10\%$ max) removed and main community. $\delta\!=\!T\!=\!$ 90 days. Motifs F1 and A6.
	Notice the model accurately tracks the change in motif frequency over time.
	}
	\label{fig:bankplotspreproc}
	\vspace{-.5cm}
\end{figure}

\xhdr{Financial Transaction Network}
In our first real-world experiment, we applied our framework to model the motif counts in a small European country's financial transaction network. The data is collected from the entire country's transaction log for all transactions larger than 50K Euros over 10 years from 2008 to 2018, and includes 118,739 nodes and 2,982,049 temporal edges. %The number of temporal edges from June 2008 to April 2008 is shown in Figure \ref{fig:bankplotsedges}.
As edges do not occur on weekends, we do not count them toward values of $T$ and $\delta$ or in computing edge rates.
Figure~\ref{fig:bankplotspreproc} compares the ground-truth motif counts and the values computed by our model with $\delta=T=$ 90 weekdays for motifs F1 and A6. Notice that \alg is able to accurately track the changes in motif counts over time.

\begin{figure}
	%\subfloat[{\label{fig:emailplots}}]{\includegraphics[width=0.42\textwidth,height=.14\textwidth]{FIG/KDDversions/email_both_00_41.png}}\\
	%\vspace{-.3cm}
	%\subfloat[{\label{fig:emailplotspreproc}}]{\includegraphics[width=0.42\textwidth,height=.14\textwidth]{FIG/KDDversions/emailthresh0p1_00_41.png}}
	\includegraphics[width=0.84\textwidth,height=.28\textwidth]{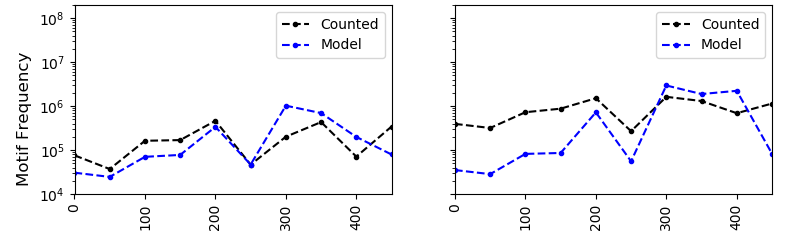}
	\vspace{-.3cm}
	\caption{Email network, $\delta\!=\!T\!=\!$ 50 days. Motifs F1 and A6. 
	Notice the model accurately tracks motif counts over time.
	%(a) full Email network (b) low-degree nodes ($<10\%$ max) removed and main community. $\delta\!=\!T\!=\!$ 50 days. Motifs F1 and A6.
	}
	\label{fig:emailplotspreproc}
	\vspace{-.5cm}
\end{figure}

\xhdr{Email Network}
We next applied our model to a network of emails exchanged within a European research institution. We use the set of 307,869 temporal edges over 977 nodes, which appear in the 500 days beginning in October of 2003. We modeled and counted frequencies of $2-$ and $3-$ node motifs with $3$ edges at a time scale of $\delta= T =$ 50 days with $10$ intervals. As shown in Figure~\ref{fig:emailplotspreproc} our model follows the trends in motif counts on the entire network. %but much more accurately models the actual values on the network formed by the largest community (selected with the same process as before).

\begin{figure}
	\includegraphics[width=0.84\textwidth,height=.32\textwidth]{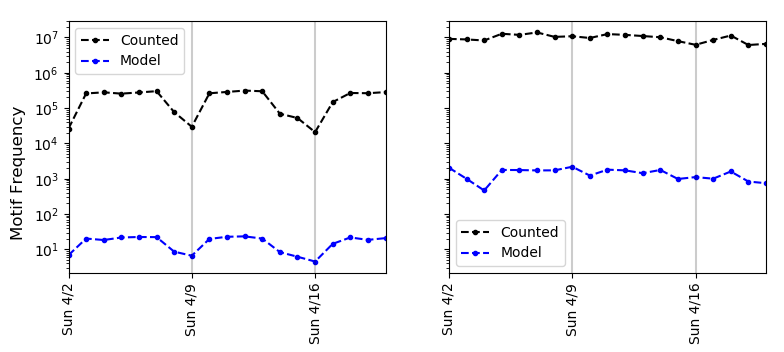}
	\centering
	\vspace{-.5cm}
	\caption{Phone call network, $\delta\!=\!T\!=\!$ 90 days. Motifs A1, A5.}
	\label{fig:phone19} 
	\vspace{-.5cm}
\end{figure}

\xhdr{Phone Call Network}
Our final dataset is a temporal network of phone calls made in April 2006 within a European country. The data includes 1,218,293 nodes and 21,907,608 temporal edges over a 19 day period. We computed the expected frequencies of $2-$ and $3-$ node motifs with $3$ edges at a time scale of $\delta= T =$ 24 hours with $19$ intervals, and $C_1=C_2=4$. Figure~\ref{fig:phone19} shows that our model accurately follows the trends in motif counts, particularly the dip in motifs around weekends, but does not fit the actual numbers of motifs. We explain this by the very localized nature of the phonecall network where both the temporal activity as well as community structure of nodes would need to be modeled. We note that our \alg is easily extendable to that setting as well.

%The community selection method did not result in modeled motif counts closer to actual motif counts on this dataset. %The plots are organized to correspond with Figure~\ref{fig:motifgrid}, so each plot corresponds to one 3-edge motif.

%\vspace{.2cm}

\subsection{Anomaly Detection}\label{sec:anomdetect}
\begin{figure}
		\subfloat[{\label{fig:syntheticanomobserve}}]{\includegraphics[width=0.48\textwidth,height=.26\textwidth]{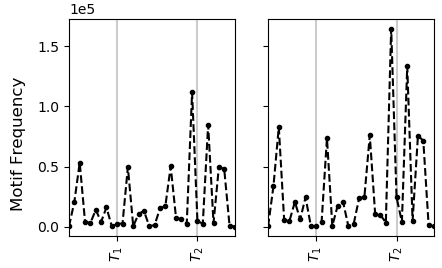}}
		 \subfloat[{\label{fig:syntheticanomdiff}}]{\includegraphics[width=0.48\textwidth,height=.26\textwidth]{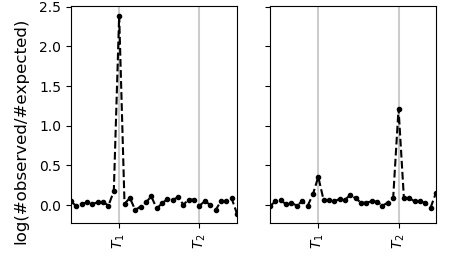}}

	\vspace{-.4cm}
	\caption{Anomaly detection in a synthetic network. (a) Motifs C3 and E2 and (b) log ratio of model and counting comparison for motifs C3 and E2. Notice the spikes in log ratio where anomalies occur.}
	\vspace{-.5cm}
\end{figure}

\xhdr{Identifying Synthetic Planted Anomalies}
We next show how our framework can be used to identify motif anomalies. 
We generate a $100$ node network over $32$ distinct intervals of $1000$ time steps with randomly chosen edge arrival rates. In addition, on the $10$-th time interval (labeled $T_1$ in Figures~\ref{fig:syntheticanomobserve} and~\ref{fig:syntheticanomdiff}) we plant anomalous reciprocated edges and on interval $25$ (labeled $T_2$) we plant anomalous repeated edges. We introduce both anomalies by the following process: after an edge is drawn from the Poisson process, we generate a corresponding anomalous edge with probability 0.25, and place it at a random time within the next $10$ and $100$ steps. %in the future.

We show that observed motif counts alone are not sufficient to identify planted motif anomalies. C3 has a reciprocated edge and E2 has a repeated edge, so the anomaly at $T_1$ should impact the number of instances of C3 while the anomaly at $T_2$ should impact the number of instances of E2. Figure~\ref{fig:syntheticanomobserve} shows the number of instances of motifs C3 and E2 over time, in which neither planted motif is noteable.
Figure~\ref{fig:syntheticanomdiff} shows the log of the ratio of counted and modeled motif frequencies. Here, we immediately see which motifs are most affected by the anomalies at $T_1$ and $T_2$.

\begin{figure}
	\subfloat[{\label{fig:bankanom}}]{\includegraphics[width=.48\textwidth,height=.28\textwidth]{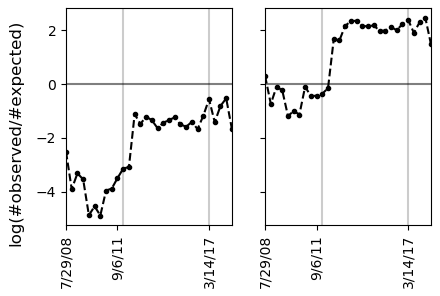}}
	\subfloat[{\label{fig:phoneanom}}]{\includegraphics[width=.48\textwidth,height=.28\textwidth]{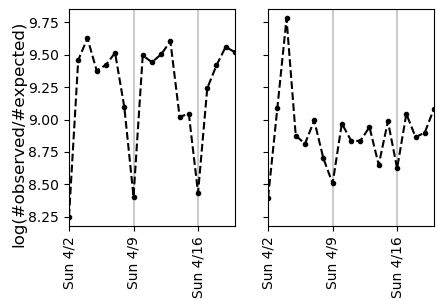}}
	\vspace{-.4cm}
	\caption{Anomaly detection in real-world networks: (a) financial transaction motifs F1 and A6 and (b) phone call motifs A1 and A5. Notice the spikes in log ratios at the times when anomalies occurred.}
	\label{fig:realdataanom}
	\vspace{-.5cm}
\end{figure}

\xhdr{Real-World Anomaly Detection}
Finally, we examine the log ratio of counted and modeled motif frequencies on real-world datasets to identify where our model most differs from the actual motif instances. Figure~\ref{fig:bankanom} shows this ratio for the financial transaction network. While the actual motif counts only go down slightly over time (Figure~\ref{fig:bankplotspreproc}) it is notable that in September of 2011, when a financial crisis hit the country, the number of modeled motifs dropped significantly, so the ratio changed as well. In Figure~\ref{fig:phoneanom} we see that the model's values for motifs A1 and A5 are more accurate on Sundays, especially A1 (a triangle).

%At approximately the time the financial crisis hits the country, the amount by which motifs are observed more than expected actually increases. In particular, while triangles are less frequent than expected before the crisis, they are about as frequent as expected during it. Other motifs are more frequent than expected throughout, but by a more significant margin during the crisis.

%\begin{figure}
%\includegraphics[width=0.48\textwidth]{FIG/casestudies/edgesovertime_bank.png}
%\centering
%\caption{Edges in bank transaction data}\label{fig:bankplotsedges} 
%\end{figure}

\section{Discussion}
\label{sec:discussion}
% !TEX root = motifs_newarxiv.tex

\hide{
We have developed an analytical null model to determine the expected number of motifs in a temporal network. Our theoretical model uses fully specified edge arrival rates and can be applied to any size motif. We also developed a simplified version of the null model given only node counts and degree distribution, such that a practical implementation is possible. Using this implementation, we first showed the accuracy of the null model by demonstrating that the expected motif counts closely match the observed counts in data generated with the same underlying model as the analytical model assumes. We also show how the null model can be useful in analyzing real-world data. The differences between observed and expected motifs allowed us to identify a change in the motif patterns in a bank transaction network during the financial crisis and in a phone call network due to weekends and the holiday Good Friday. In the future, it would be useful to look at ways to simplify the null model for practicality other than using a block model.
}

We have developed an analytical model to determine the expected number as well as the variance of motifs in a temporal network. We developed an efficient parameter inference technique as well as provided closed form solutions for the expected motif frequencies in the general case where temporal edges appear with distinct rates between different pairs of nodes, and the arrival rate of temporal edges between every pair of nodes may change over time. %Furthermore, we proposed a stochastic block model by grouping the nodes based on their levels of activity, and provided accurate estimations from the expected motif frequencies. %In our experiments, we show that our framework can provide accurate estimations from the expected motif frequencies based on the block model. A notable advantage of our analytical model is that unlike previous methods, our framework does not require generating randomized network ensembles, and thus scales well to large dynamic networks.
We demonstrated the effectiveness of our Temporal Activity State Block Model combined with our analytical model of temporal motifs for modeling temporal networks and detecting anomalies. Applied to a financial transaction network, our framework can successfully 
localize anomalies caused by a financial crisis. Moreover, we identify trends such as weekends by looking at the significance profile of temporal motifs in a phone call network. 

%at different time scales.
%In the future, it would be useful to look at ways to simplify the null model for practicality other than using a block model.

%\section{Conclusion}
%\label{sec:conclusion}
%\input{060conclusion}

\bibliography{refs}
\bibliographystyle{abbrv}

\appendix
\section{Appendix }
\label{sec:appendix}

\subsection{Additional Figures}
\begin{figure}[H]
	\centering
	\vspace{-4mm}
	\subfloat[{}]{\includegraphics[width=.26\textwidth]{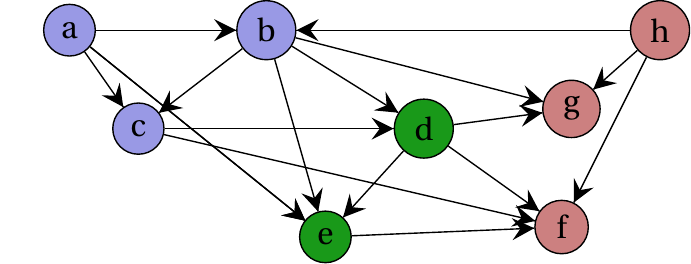}} 
	\subfloat[{}]{\includegraphics[width=.22\textwidth]{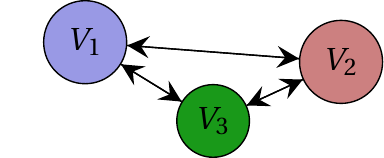}} 
	\\\vspace{-2mm}
		\subfloat[\label{fig:assignment}]{\includegraphics[width=.18\textwidth]{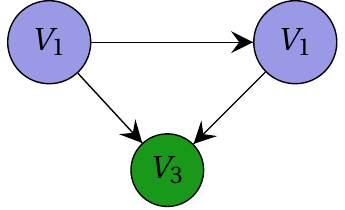}} 
		\subfloat[\label{fig:bij1}]{\includegraphics[width=.3\textwidth]{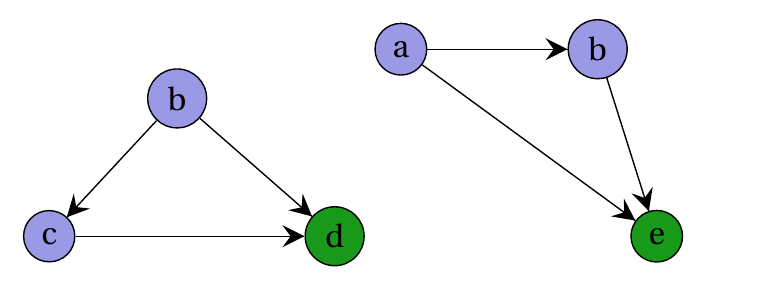}} 
	\vspace{-2mm}
	\caption{(a) Example graph, (b) assignment of partition $V= \{V_1,V_2,V_3\}$, (c) activity state assignment to a motif, and (d) two instances of motifs with the assigned activity states. }
	\label{fig:isomorphic}
	\vspace{-.1cm}
\end{figure}

Here we provide additional figures explaining our analytical model. Figure~\ref{fig:var} gives examples of joint instances of motifs, used in computing variance. Figure~\ref{fig:isomorphic} shows how multiple distinct motif instances may correspond to the same activity state assignment, a key feature in reducing the complexity of our model.

\begin{figure}[H]
	\includegraphics[width=0.48\textwidth]{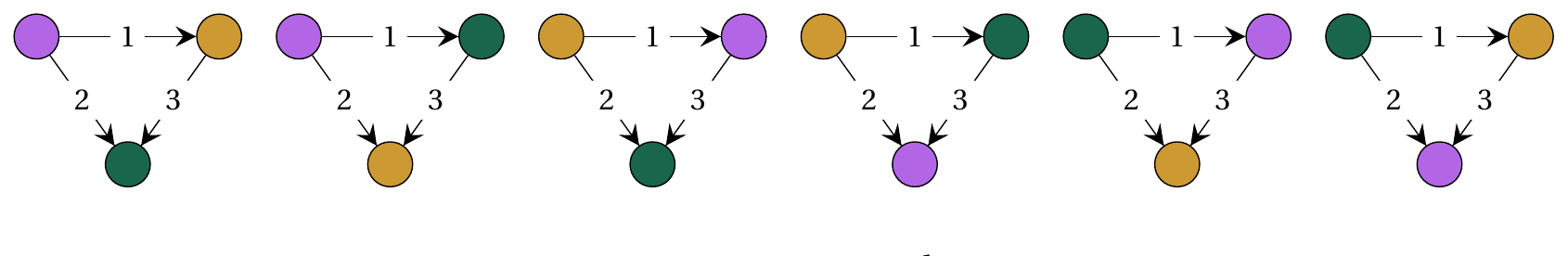}
	\centering
	\vspace{-.6cm}
	\caption{$3!=6$ unique bijections between a $3$-node $3$-edge subgraph and the motif in Figure \ref{fig:1b}. Some of these bijections correspond to the same static graph, but the ordering on the temporal motif edges are different.
	}\label{fig:permutationex} 
	\vspace{-.4cm}
\end{figure}

\begin{figure}[H]
	\centering
	\subfloat[\label{fig:shared1}]{\includegraphics[width=.25\textwidth]{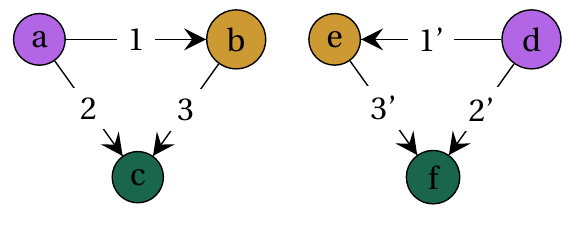}} 
		\subfloat[\label{fig:shared2}]{\includegraphics[width=.21\textwidth]{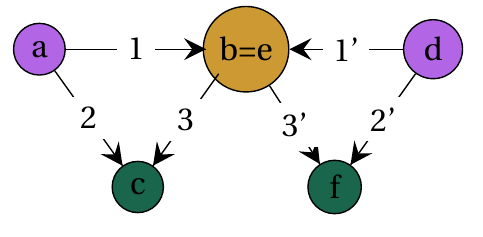}} 
			\\\vspace{-3mm}
		\subfloat[\label{fig:shared3}]{\includegraphics[width=.18\textwidth]{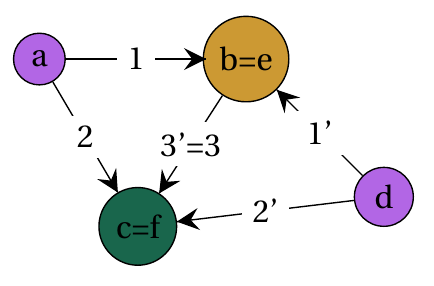}} 
		\subfloat[\label{fig:shared4}]{\includegraphics[width=.14\textwidth]{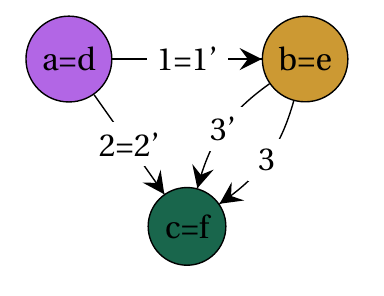}} 
	\vspace{-2mm}
	\caption{Examples of joint instances of motif $M$: $S_1$ on vertices $(a,b,c)$ with edges labeled $1,2$, and $3$ at times $t_1<t_2<t_3$ and $S_2$ on vertices $(d,e,f)$ with edges labeled $1',2'$, and $3'$ at times $t_{1'}<t_{2'}<t_{3'}$. }
	\label{fig:var}
	\vspace{-.4cm}
\end{figure}

 \label{fig:1}

\subsection{Implementation Details}
Our C++ implementation of the model can be found at:\\
\url{https://www.dropbox.com/sh/81tkcf1amemmq2i/AAC2k1vdX2DhsbFQz9LCzovEa?dl=0}
\\Also included is the set of synthetic graph generators and the setup for comparing our model's run time to motif counting on randomly generated graphs. We use the implementation of temporal motif counting from~\cite{paranjape2017motifs} included in the Snap library, available at:
\url{http://snap.stanford.edu/snap/index.html}
\\The European research institute email network dataset is publicly available and can be found at:\\
\url{http://snap.stanford.edu/data/email-EuAll.html}

We pre-processed all datasets by sorting by timestamp and relabeling nodes as integers appearing in increasing order; the derived sets from the email dataset can also be found with our code. We also constructed a further processed graph by removing nodes with degree less than 10\% the maximum degree, and then selecting the subgraph induced by the remaining nodes forming the largest community. 

 Table~\ref{table:runtime} shows the timings results of our model and for counting motifs over randomly generated graphs.  The average motif counts in synthetic graphs generated according to the activity state block model will converge to the values generated by our analytical model. For each of the two methods, we measured the run time on average over 100 trials. For the analytical model a trial consists of generated expected motif values from the \alg state groups and edge arrival rates. State groups and edge arrival rates for all timing experiments were derived from fitting the model on the financial transaction network. There are approximately 13,000 nodes and the number of edges in each experiment ranged from about 200 in the shortest time interval to approximately 1.8 million in the longest. For the random ensemble method, a trial consists of generating a random graph according to the activity state blocks in the \alg and then counting motif instances. In practice, the random graph counting method would need to run multiple trials and average the resulting motif counts to achieve approximately the same values as the analytical model, so the run time would be even greater than shown here, e.g. 100 times longer to get the average motif counts over an ensemble of 100 random graphs.

\begin{table}[!t]%\vspace{-.3cm}
	\small
	\centering
	\begin{center}\vspace{-.2cm}
		\begin{tabular}{||c c c ||} 
			\hline
			Time Interval & Analytical Model & Random Graph Counting \\\hline \hline
			1e3  & 6.40 & 10.2 \\\hline
			1e5 & 6.50 & 10.6 \\\hline
			1e7&  6.45& 164 \\ 
					\hline
		\end{tabular}
	\end{center}
	\caption{Average run times in seconds for analytical model and randomized ensemble of 100 graphs. Input to each test is the same set of 36 activity state group sizes with corresponding rates over increasing time intervals. }\label{table:runtime}
\end{table}

\subsection{Additional Experiments}
\begin{figure*}[t]
	\centering
		\subfloat[{}]{\includegraphics[width=.33\textwidth,valign=t]{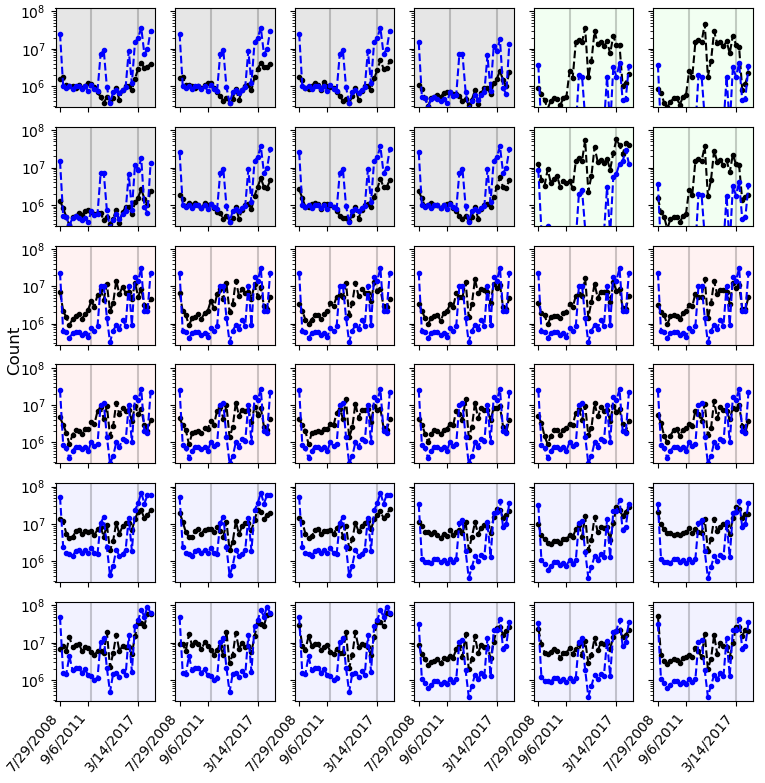}} 
	\subfloat[{}]{\includegraphics[width=.33\textwidth,valign=t]{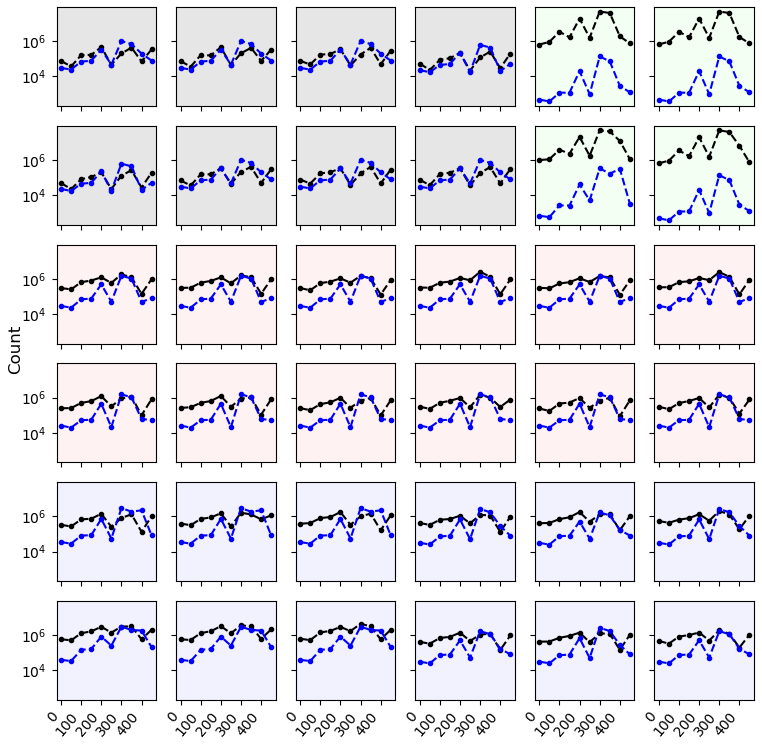}} 
		\subfloat[\label{fig:assignment}]{\includegraphics[width=.33\textwidth,valign=t]{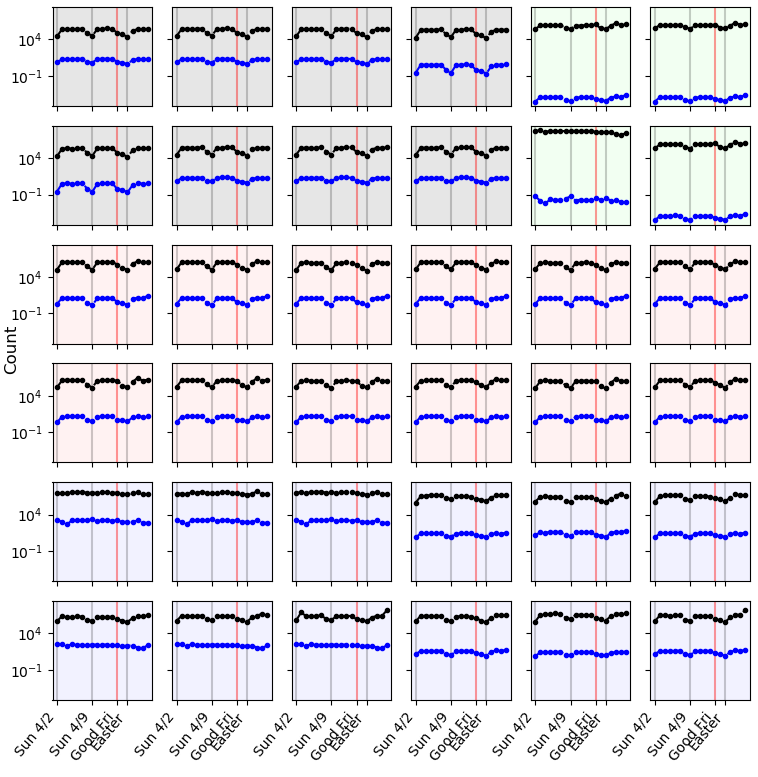}} 
	
	\caption{Counted (black) and \alg (blue) motif results on all 36 motifs: (a) financial transaction network (b) email network and (c) phone call network. }
	\label{fig:plottrio}
\end{figure*}

\begin{figure*}[t]
	\centering
				\subfloat[\label{fig:assignment}]{\includegraphics[width=.33\textwidth,valign=t]{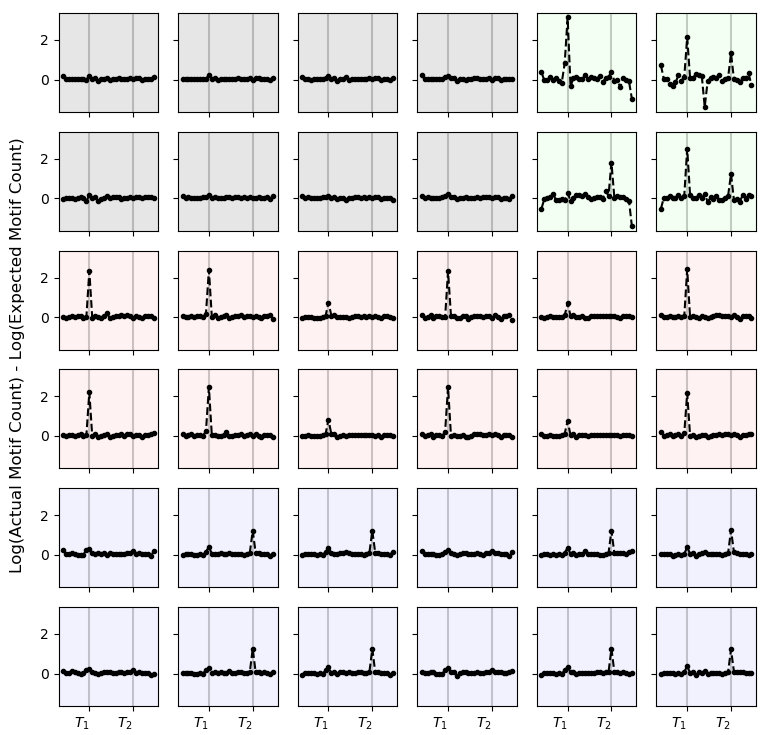}} 
		\subfloat[\label{fig:logratio}]{\includegraphics[width=.33\textwidth, valign=t]{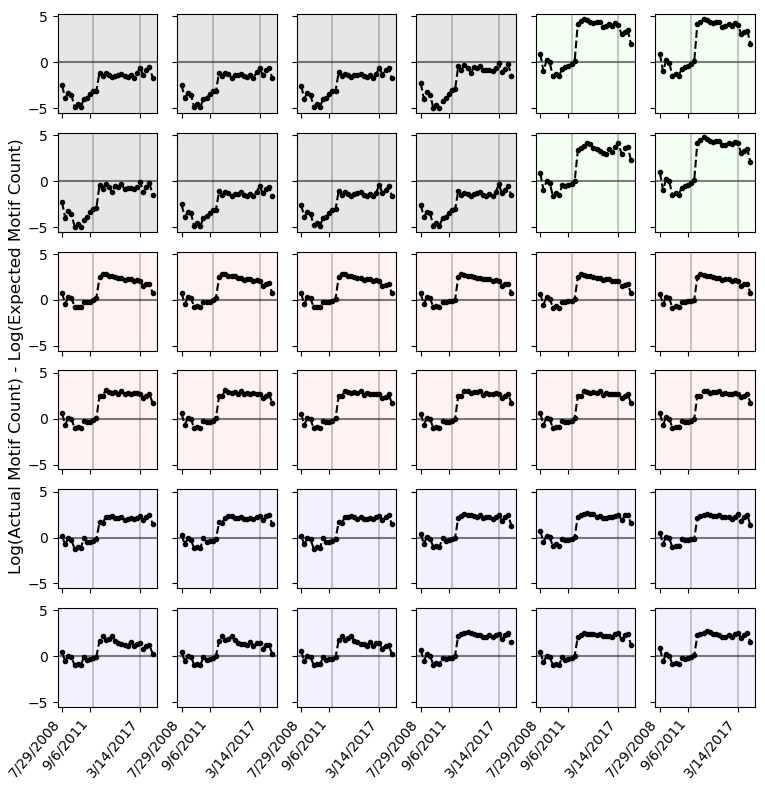}} 
	\subfloat[\label{fig:logration}]{\includegraphics[width=.33\textwidth,valign=t]{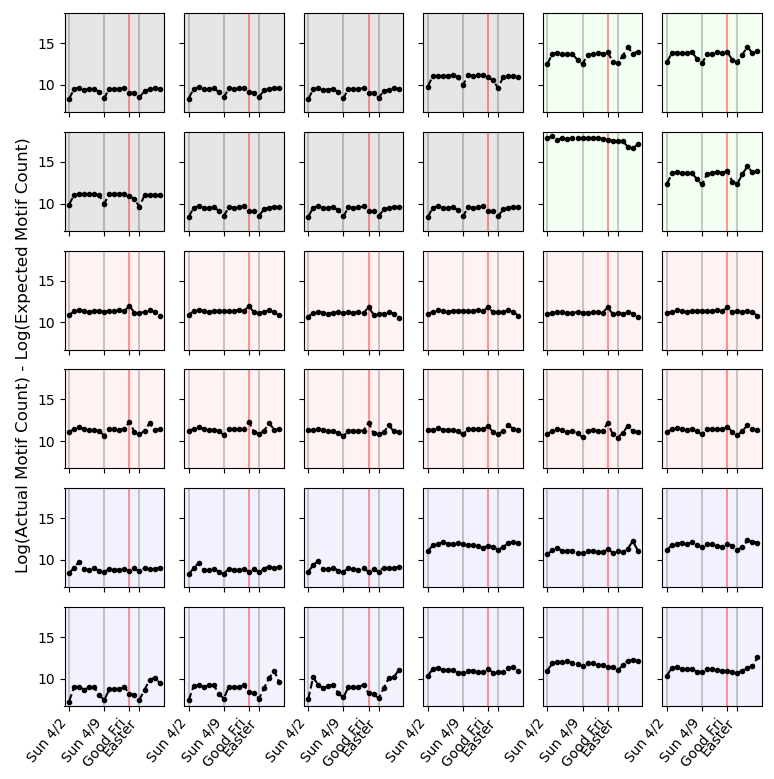}} 
		
	\caption{Logarithm of ratio between counted and \alg motif instances for (a) synthetic network with planted anomalies (b) financial transaction network and (c) phone call network. }
	\label{fig:plottrio2}
\end{figure*}

Here we show the result of our experiments on the full set of 36 motifs. Plots are organized corresponding to Figure~\ref{fig:motifgrid}. Figure~\ref{fig:plottrio} includes the experiments shown in Section~\ref{sec:realworldresults} and Figure~\ref{fig:plottrio2} includes the experiments show in Section~\ref{sec:anomdetect}.


\end{document}